\documentclass[10pt,a4paper]{article}
\usepackage{amsmath}
\usepackage{amssymb}
\usepackage{graphicx}

\usepackage{cite}

\usepackage{color} 

\newcommand{\pmm}{p_{\text{m}}}
\newcommand{\pii}{p_{\text{i}}}
\newcommand{\nc}{N_{\text{c}}}

\newcommand{\bnc}{\bar{N}_{\text{c}}}
\newcommand{\bnt}{\bar{N}_{\text{t}}}
\newcommand{\be}{\begin{equation}}
\newcommand{\ee}{\end{equation}}
\newcommand{\nofigure}{0} %%mettere 0 per avere figure, 1 per non averle

\makeatletter
\renewcommand{\@makecaption}[2]{
   \vskip\abovecaptionskip
   \sbox\@tempboxa{#1. #2}%
   \ifdim \wd\@tempboxa >\hsize
      \hspace*{1cm}\begin{minipage}{14cm}{\bf #1.} #2\end{minipage}%\par
   \else
     \global \@minipagefalse
     \hb@xt@\hsize{\hfil\box\@tempboxa\hfil}%
   \fi
   \vskip\belowcaptionskip}
\makeatother

\begin{document}

\begin{center}{\huge \textbf{Mutation-selection dynamics and error threshold in an evolutionary
model for Turing Machines}}

\vspace{10mm}
\begin{minipage}[]{110mm}\begin{center}
{\Large Fabio Musso$\mbox{}^{\S}$ \& Giovanni Feverati$\mbox{}^{\dag}$}\\[10mm]
$\mbox{}^{\S}$ Departamento de F\'isica, Universidad de Burgos,\\ 
Plaza Misael Ba\~{n}uelos s/n, 09001 Burgos, Spain\\
fmusso@ubu.es\\[2mm]
$\mbox{}^{\dag}$ Laboratoire de physique theorique LAPTH, CNRS, Universit\'e de Savoie,\\ 
9, Chemin de Bellevue, BP 110, 74941, Annecy le Vieux Cedex, France
feverati@lapp.in2p3.fr
\end{center}
\end{minipage}
\end{center}

%%%%%%%%%%%%keywords

\section*{Abstract}
\begin{quote}
%%%%%%%\textbf{Background}
We investigate the mutation-selection dynamics for an evolutionary computation model based on Turing Machines that we introduced in a previous article \cite{PRE}.  

%%%%%%\textbf{Methodology/Principal findings}
The use of Turing Machines allows for very simple mechanisms of code growth and code activation/inactivation through point mutations.
To any value of the point mutation probability corresponds a maximum amount of active code that can be maintained by selection and the Turing machines 
that reach it are said to be at the error threshold. Simulations with our model show that the Turing machines population evolve towards the error threshold.

%%%%%\textbf{Conclusions/Significance}
Mathematical descriptions of the model point out that this behaviour is due more to the mutation-selection dynamics than to the intrinsic nature of the Turing machines.
This indicates that this result is much more general than the model considered here and could play a role also in biological evolution. 
\end{quote}

%%%%%%% acknowledgments

%Spanish Ministerio de Ciencia e Innovacion under grant MTM2007-67389 (with EU-FEDER support), by
%Junta de Castilla y Leon (Project GR224) and by UBU-Caja de Burgos(Project K07J0I)

% Please keep the Author Summary between 150 and 200 words
% Use first person. PLoS ONE authors please skip this step. 
% Author Summary not valid for PLoS ONE submissions.   
%\section*{Author Summary}
%Despite an increasing interest in the study of ``in silico'' evolutionary models, many biologists remain skeptical on their effective usefulness. The most common objection is that 
%even the most refined simulation is unable to reproduce the details and the complexity of a real biological organism and, as a consequence, digital evolution models are inadequate tools
%for the study of biological evolution. However, some of the phenomena observed could be due to the mutation-selection dynamics more than to the specific nature of the evolving objects. 
%In these cases, the results have a much wider degree of generality than the specific model under consideration. We show that this is what happens in our evolutionary model, 
%where the mutation-selection dynamics pushes the Turing machines toward the error threshold (the maximum mutation probability beyond which selection is unable to preserve the informative
%content of the genomes).      

\section{Introduction}

The study of ``in silico'' evolutionary models has increased significantly in recent times, see \cite{Tierra}, \cite{Lenskietal1999}, \cite{Wilkeetal2001}, \cite{Lenskietal2003}, \cite{Knibbeetal2007}, \cite{Knibbeetal2007b}, \cite{Cluneetal2008}, \cite{PRE} just to give some examples. The basic idea behind these models is to simulate the evolution of computer algorithms 
subject to mutation and selection procedures. In this artificial evolution setting, the algorithms play the role of the biological organisms and they are selected on the basis of their 
ability in performing one or more prescribed tasks (replicate themselves, compute some mathematical function, etc.). 
While the simulated algorithms have clearly an incomparably lesser degree of complexity than a whatever biological organism, the hope is that (at least some of) the phenomena observed in the digital evolution model could correspond to general behaviours of evolutionary systems. Indeed, it seems that this is what happens in some cases:
emergence of parasitism in \cite{Tierra}, quasi-species selection in \cite{Wilkeetal2001} and the striking similarity between the C-value enigma \cite{Gregory} and the phenomenon
of code-bloat in evolutionary programming \cite{Luke}, \cite{PRE}. 

One of the motivations for performing artificial evolution experiments is the continuously increasing computational power of modern computers. Nowadays, very fast multiprocessor
computers have relatively low prices and many scientific institutions have at their disposal large facilities for parallel computation. For example, one run lasting 
$50000$ generations of a population of $300$ Turing machines (TMs) of our evolutionary model lasts about half a day per processor on an ordinary home computer  (for the higher value of 
the states-increase rate $\pii$, for 
lower values it lasts considerably less). The long term evolution experiment on E. coli directed by R.E. Lenski reached the $40000$ generations after almost $20$ years \cite{Lenski} 
(however, the population considered in this experiment is much larger, of the order of $10^7$ cells). When population size is not a crucial parameter, digital evolution experiments     
can explore a number of generations inaccessible to laboratory experiments with real organisms.  If one wants to study evolutionary effects
on a so large time scale in real biological organisms, then has to resort to paleontological studies. However, such studies are vexed by the incompleteness of the fossil record
and by the unrepeatability of the experiments. Indeed, repeatability allows to discriminate easily among effects due to adaptation and those simply due to drift. 
These problems are overcame in laboratory experiments such as Lenski one, but at the price of reducing the environment to a Petri dish. Artificial evolution experiments allow to 
explore larger time scales than laboratory experiments at much reduced costs, but at the higher price of replacing biological organisms with algorithms. 
By the way, there is another big advantage when performing artificial evolution experiments, namely the complete control over all the experimental settings.  
This gives the opportunity to use a reductionistic approach, by studying separately the effects of the various mechanisms involved in the evolutionary dynamics, something that is 
very difficult to obtain when working with real organisms. Finally, as a last argument in favour of artificial evolution experiments, we cite one given by Maynard Smith 
\cite{MaynardSmith}: ``...we badly need a comparative biology. So far, we have been able to study only one evolving system and we cannot wait for interstellar flight to provide us 
with a second. If we want to discover generalizations about evolving systems, we will have to look at artificial ones.''

As we said, even the most complicated computer algorithm is incomparably simpler than a whatever biological organism. Moreover, typical artificial evolution experiments have a unique 
ecological niche and the interaction between the artificial organisms is often limited to the comparison of their performances. So, a very big distance separates
artificial evolution experiments from  biological evolution. For this reason, many biologists are skeptical on the biological relevance of the results obtained in the digital 
framework; for example, some objections typically raised are reported in \cite{ONeill}. On the other hand, supporters of artificial evolution experiments reply that the observed
results can actually be general phenomena of evolutive systems, therefore being independent from the particular model under consideration. To test this hypothesis it would be nice
to compare the results obtained in the artificial evolution setting with real biological data, but this is very hard to do for long-term evolutionary effects, that is where artificial evolution models are most useful.  
On the other hand, general evolutionary behaviours do emerge if the mutation-selection dynamics have a
prominent role on the peculiar characteristics of the evolving organism. When this is the case, 
the observed effects can be reproduced through a population genetic mathematical model. Indeed, these models center on the dynamics induced by the selection and mutation
operators (under some work hypotheses), more than in the specific details of functioning of the organism. If successful, this procedure extends the validity of the results observed in the evolutionary model under consideration to all 
the evolutionary models working with the same mutation and selection operators (under the same hypotheses). 
This means that the problem of the biological relevance of the results obtained in the artificial evolution experiment is switched to 
the problem of assessing the biological likelihood of the mutation and selection operators and of the hypotheses used in the mathematical model.    

In this paper we apply this strategy, so that we derive a deterministic and a stochastic population genetic model of our evolutionary model for TMs. Our main aim is to show that
the evolutionary dynamics pushes the TMs toward the error threshold \cite{Eigen71}, 
%\textcolor{blue}{namely the point where the population degrades due to bad mutations faster than
%it improves by selection of the fittest.}
The population genetic model is used to compute mathematically the value of the error
threshold and to show that this dynamical behaviour is due to quite mild hypotheses
%\textcolor{blue}{that often are realized in the biological context.}

According to this program, in the Materials and Methods section we first briefly recall our evolutionary model for TMs \cite{PRE} and the Eigen error threshold concept.
A deterministic population genetic model for our digital evolution model is introduced in the third subsection ``The deterministic model'', while its stochastic counterpart (limited to the evolution of the 
best performing TMs) is given in the fourth one. In the Results section we report the results obtained by the computer simulations  and compare them with 
those predicted by the mathematical models. 
%\textcolor{blue}{Here and in the Discussion section, we give also the biological motivations of the 
%model features.}
Finally, our concluding remarks are given in the Discussion section.

\section{Materials and Methods}

\subsection{The evolutionary model} \label{model}

We basically use the same evolutionary programming model based on Turing Machines that has been introduced in \cite{PRE}. The following are the only differences between
that model and the model we use in this article:
\begin{enumerate}
\item the TMs' movable head can move only right or left, now it cannot stay still (this also affects the definition of the added state); \label{still}
\item the TMs' tape is now circular, so that the TM head cannot exit from the tape; \label{circular}
\end{enumerate}  
The first choice allows us to save one bit of memory for each state of the TMs and, at the same time, makes our definition more similar to the original one
\cite{Turing}. The second choice seems to us the most convenient when dealing with finite tapes.
  
To the sake of making this article self-contained, we give a terse description of Turing machines and of the evolutionary programming model that we use.

Turing Machines are very simple symbol-manipulating devices which can be used to encode any feasible algorithm. 
They were invented in 1936 by Alan Turing \cite{Turing} and used as abstract tools to investigate the problem
of functions computability. For a complete treatment of this subject we refer to  \cite{davis}.

\begin{figure}[!ht]
\ifcase\nofigure
  \includegraphics{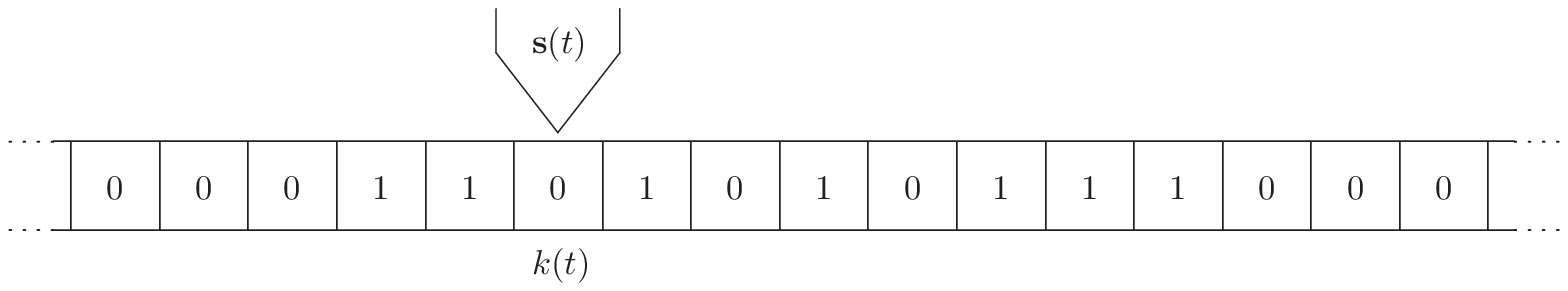}
\or
\fi
\caption{{\bf Graphical representation of a Turing machine.} The machine is shown at time $t$, in the internal state 
$\mathbf{s}(t)$, located on the $k(t)$-th cell of a infinite tape.\label{Turing}}
\end{figure}

A Turing machine consists of a movable head acting on an infinite tape $T(t)$, see figure \ref{Turing}. The tape consists of discrete cells that can contain
a 0 or a 1 symbol. The head has a finite number of internal states that we denote by ${\bf N}$ (in which case the TM is called an ${\bf N}$-state TM).
At any time $t$ the head is in a given internal state $\mathbf{s}(t)$ and it is located upon a single cell $k(t)$ of the infinite tape $T(t)$.
It reads the symbol stored inside the cell and, according to its internal state and the symbol read, performs
three actions: 
\begin{enumerate}
\item ``write'':  writes a new symbol on the $k(t)$ cell ($T(t) \mapsto T(t+1)$), 
\item ``move'': moves one cell on the right or on the left ($k(t) \mapsto k(t+1)$),
\item ``call'': changes its internal state to a new state ($\mathbf{s}(t) \mapsto \mathbf{s}(t+1)$).
\end{enumerate}       
Accordingly, a state can be specified by two triplets ``write-move-call'' listing the actions to undertake 
after reading respectively a $0$ or $1$ symbol. There exists a distinguished state (the Halt state) that stops 
the machine when called. The initial tape $T(0)$ is the input tape of the TM, and the tape $T(\bar{t})$ at the instant $\bar{t}$
when the machine stops is its output tape, that is the result of executing the algorithm defined by the given TM on the input tape $T(0)$. 
However, many TMs will never stop, so that they will not be associated with any algorithm. Moreover, the halting problem, that is the problem 
of establishing if a TM will eventually stop when provided with a given input tape, is undecidable. This means that there will exist TMs for which it is
impossible to predict if they will eventually halt or not for a given input tape.

We have to introduce some restrictions on the definition of the TMs in our evolutionary model.     
Since we want to perform computer simulations, we need to use a tape of finite length that we fix to $300$ cells. 
The position of the head is taken modulo the length of the tape,
that is we consider a circular tape with cell $1$ coming next cell $300$. 
Since it is quite easy to generate machines that run forever, we also need to fix a maximum number of time steps, 
therefore we choose to force halting the machine if it reaches $4000$ steps.

We begin with a population of $300$ $1$-state TMs of the following form 
\begin{equation}
\begin{array}{|c|c|}
\hline
 &\bf 1 \\ \hline
 0 & 0-\mbox{move1}-\mbox{\bf Halt} \\ \hline
 1 & 1-\mbox{move2}-\mbox{\bf Halt} \\ \hline
\end{array} \label{state}
\end{equation}
where move1 and move2 are fixed at random as Right or Left,
and let them evolve for $50000$ generations.
At each generation every TM undergoes the following three processes (in this order):
\begin{enumerate}
\item states-increase,
\item mutation,
\item selection and reproduction.
\end{enumerate}

\paragraph{States-increase.} In this phase, further states are added to the TM with a rate $\pii$. The new states are the same as (\ref{state}) with
the ${\bf 1}$ label replaced by ${\bf N+1}$, ${\bf N}$ being the number of states before the addition.
%\footnote{
While it is clear that the states-increase should be considered a form of mutation (vaguely resembling insertion), we preferred to keep it distinguished because its effect is always neutral. %}  

\paragraph{Mutation.}
During mutation, all entries of each state of the TM are randomly changed with probability $\pmm$. The new entry is randomly chosen 
among all corresponding permitted values excluding the original one. 
The permitted values are: 
\begin{itemize}
\item 0 or 1 for the ``write'' entries;
\item Right, Left for the ``move'' entries;
\item The Halt state or an integer from {\bf 1} to the number of states $\mathbf N$ of the machine for the ``call'' entries.
\end{itemize}

\paragraph{Selection and reproduction.}
In the selection and reproduction phase a new population is created from the actual one (old population).
The number of offspring of a TM is determined by its ``performance'' and, to a minor extent, by chance.
%\footnote{
Actually, in the field of evolutionary programming the word used is fitness. 
However, in population genetics, fitness is used to denote the expected number of offspring (or the fraction that reach the reproductive age) of an individual. To avoid ambiguities, we decided to reserve the word fitness for this meaning, and to use the word performance for the evolutionary programming one. %} 
The performance of a TM is a function that measures how well the output tape of the machine reproduces a given ``goal'' tape
starting from a prescribed input tape. We compute it in the following way. The performance is initially set to zero. 
Then the output tape and the goal tape are compared cell by cell. The performance is increased by one for any $1$ on the
output tape that has a matching $1$ on the goal tape and it is decreased by 3 for any $1$ on the output tape
that matches a $0$ on the goal tape. 
    
As a selection process, we use what in the field of evolutionary algorithms is known as ``tournament selection of size 2 without replacement''. 
Namely two TMs are randomly extracted from the old population, they run on the input tape and a performance value is
assigned to them according to their output tapes. The performance values are compared and the machine which scores higher creates two copies 
of itself in the new population, while the other one is eliminated (asexual reproduction). If the performance values are equal, each TM creates
a copy of itself in the new population.
The two TMs that were chosen for the tournament are eliminated from the old population (namely they are not replaced) and the process restarts until the exhaustion 
of the old population. Notice that this selection procedure keeps the total population size $N$ (with $N$ an even number) constant.
From our point of view this selection mechanism has two main advantages: it is computationally fast and quite simple to treat mathematically.

The choice of TMs to encode the algorithms in our evolutionary model was convenient for various reasons. The first reason is that any feasible algorithm can be encoded 
through a TM (Church-Turing thesis \cite{davis}); so that TMs are universal objects inside the algorithms class. The second reason is that even TMs with a very low number of states can 
exhibit a very complicated and unpredictable behaviour (even if the input tape is filled only with zeroes as in our case, see for example the busy beaver function \cite{Beaver}).
Thanks to this property, it is very difficult to predict the dynamics of our evolutionary model.

While developing the model, we were primarily interested in how the variations in the length of the code affect the evolutionary dynamics. 
From this point of view, the TMs present many advantages. The distinction between coding and non-coding triplets is unambiguous and very easy to verify. We define a triplet
as non-coding (with respect to a given input tape) if mutations in its entries cannot affect the output tape of the TM and we will call it coding in the complementary case.
This definition is practically equivalent to saying that a triplet is coding if it is executed at least one time when the TM runs on a given input tape and it is non-coding if it is never executed.
In this way, to identify coding triplets, one has only to run the TM on the prescribed input tape and mark the triplets that are executed. Another advantage is that the mechanism
of state-adding is completely neutral; the added states are always non-coding, so that they cannot change the performance of the TM. On the other hand, there is a simple mechanism of code activation.
Namely, a triplet of a non-coding state $s$ can be activated, for example, when a mutation occurs in the call entry of a coding state changing its value to $s$, but notice that also mutations in 
the write and move entries (of a coding triplet), can result in an activation or inactivation of the TMs triplets.	
Finally, another advantage of using TMs is that they are specified in terms of an atomic instruction: the state. 

\subsection{The Eigen's error threshold} \label{error}

The error threshold concept was introduced in 1971 by Eigen  in the context of its quasispecies model \cite{Eigen71},\cite{Eigen-Schuster}. The model describes the 
dynamics of a population of self-replicating polynucleotides of fixed length $L$, subject to mutation and under the constraint of constant population size.
Each polynucleotide $I^{(i)}$ is characterized by its replication rate $A_i$, its degradation rate $D_i$ and the probabilities $Q_{ji}$ of mutating into a different
polynucleotide $I^{(j)}$ as a consequence of an inexact replication. All these parameters are assumed to be fixed numbers, independent of time and of population composition.
The Eigen model then consists of a set of ODEs determining the evolution of the frequency $\phi_i$ of the polynucleotides $I^{(i)}$ in the total population:
\begin{equation}
\dot{\phi_i}=\sum_j (A_j Q_{ij}-D_j \delta_{ij}) \phi_j  -  \phi_i \sum_j (A_j - D_j) \phi_j, \label{Eigen}
\end{equation}
where the sum is over all possible polynucleotide templates $I^{(j)}$. It is supposed that the polynucleotide $I^{(1)}$ has a larger fitness than the others
$A_1-D_1> A_k-D_k, \ k>1$. Such polynucleotide is usually called the master sequence while the others are called mutants. 
If we assume that mutation is exclusively due to point mutation, we neglect transversions, suppose that transitions have all the same probabilities of occurring and that the point mutation probability is independent on the site, then we can identify our polynucleotides as binary chains of length $L$ and the mutation probabilities 
$Q_{ji}$ depend only on the point mutation probability $q$ and the Hamming distance $d(i,j)$ among the binary chain $I^{(i)}$ and the binary chain $I^{(j)}$:
$$
Q_{ji}=q^{d(i,j)}(1-q)^{L-d(i,j)}
$$      
Once assigned the $A_j$ and $D_j$ parameters, one can study the asymptotic composition of the population as a function of the point mutation probability $q$. 
It turns out that (at least for some choices of the fitness landscape, see \cite{Swetina-Schuster}, \cite{Krall}, \cite{Wilke}, \cite{Takeuchi}, \cite{Saakiaan}) there is a sharp transition in the population composition near a particular value of $q$ that is termed 
error threshold. Before the error threshold, the population is organized as a cloud of mutants surrounding the master sequence, while,
after the error threshold, each polynucleotide is almost equally represented. In the thermodynamic limit (when the chain length $L$ goes to infinity and the point mutation
$q$ goes to zero in such a way that the genomic mutation rate $p=qL$ stays finite)
this is a real phase transition of first order \cite{Tarazona}, and the error threshold is mathematically well defined. As a consequence, from this model it follows that natural selection can preserve the genome informative content  only if the mutation rate is lower than the error threshold (see \cite{Eigen2000}); after the error threshold, all the information content is lost.
For a single peak fitness landscape (i.e. $A_k-D_k=A_2-D_2, \ k>2$) and in the thermodynamic limit, the system of equations (\ref{Eigen}) can be decoupled into a two by two system by introducing a 
collective variable $\phi_M$ for the overall frequency of mutants in the population
$$
\phi_M= \sum_{k=2}^{\infty} \phi_k
$$ 
In the thermodynamic limit, the fidelity rate of the master sequence will be given by $Q_{11}=e^{-p}$ and the probability of back mutation $Q_{1M}$ will go to zero. So, the Eigen equations take the form: 
\begin{eqnarray}
\dot{\phi}_1& =& \left(A_1 e^{-p}-D_1 \right) \phi_1 - \phi_1 \left[ \left( A_1-D_1 \right)\phi_1 + \left( A_2-D_2 \right) \phi_M \right] , \label{Eigen1}\\
\dot{\phi}_M& =& A_1 \left( 1-e^{-p} \right) \phi_1  + (A_2 - D_2) \phi_M-\phi_M  \left[ \left( A_1-D_1 \right)\phi_1 + \left( A_2-D_2 \right) \phi_M \right].  \label{Eigen2}
\end{eqnarray} 
The error threshold $\bar{P}$, in this case, will coincide with the lowest value of the mutation probability $P=1-e^{-p}$ for which the master sequence goes extinct in the asymptotic limit $t \to \infty$ \cite{Nowak}. Using the constraint $\phi_1+ \phi_M=1$ in equation (\ref{Eigen1}) we get a closed equation for $\phi_1$ that gives:  
\begin{equation}
\bar{P}=\frac{(A_1-D_1)-(A_2-D_2)}{A_1}. \label{Pt}
\end{equation}

Observe that the infinite population limit has the effect of removing the genetic drift and that the survival of the master sequence in the asymptotic limit for values of the mutation 
probability less than the error threshold is possible only for infinite populations. For finite populations, when the probability of reverse mutations is zero, the genetic drift will always
push the population in its only absorbing state: the extinction of the master sequence. In the finite population case, however, the expected number of generations before the extinction of the master sequence will start to grow by several orders of magnitude when the mutation probability drops below the error threshold (see \cite{Nowak89}, \cite{io}).
%\footnote{
This is the reason why 
the value of the error threshold predicted by the deterministic model works also for finite populations.%} 
This effect is also present in our model (see the next two sections and figure \ref{ext1}).

\subsection{The deterministic model} %\label{deterministic}

In this section we will describe a deterministic mutation-selection model with tournament selection of rank two and we will obtain the corresponding error threshold. 
Our model of selection and reproduction is very different from that of the Eigen model. In particular, the number of offspring is not constant for each genotype, since while the performance landscape is fixed, the fitness landscape 
changes in time following the changes in the population composition.
Despite this fact, when one neglects the probability of back mutations, one obtains a closed equation for the number of individuals with the best performance, as it happens for the Eigen model with the single peak fitness landscape. 
Following the example of the Eigen model (see also \cite{bull2005}), we will define the error threshold as the value of the mutation probability that causes the extinction of the master
sequence (that, in our case, is the best performance class) for our selection model considered in the deterministic limit. 

Let us suppose that we have $M$ possible performance classes, a population of size $N$, and let us denote with $n_i$ the number of individuals belonging to the $i$th performance class. 
In the selection step we draw $2$ individuals from the population without replacement and compare their performances.
The individual with higher performance is copied into the new population and has a probability $f$ to give raise to another copy, while, with probability $1-f$ the second copy will
belong to the individual with the lower performance. When the two individuals have the same performance, then both are passed to the new population. The two individuals
are eliminated from the old population and the process is restarted until the old population is exhausted and the new population is replenished. 
Notice that it must hold $0 \leq f \leq 1$. The selection mechanism we used in our TMs model correspond to the particular choice $f=1$.

With this mechanism, each individual belonging to $n_i$ has a probability
$$
P_{2}=f \frac{\sum_{j<i} n_j}{N-1}
$$
of making two copies of itself, a probability
$$
P_1=\frac{1}{N-1} \left( (1-f) \sum_{j<i} n_j + n_i-1+ (1-f) \sum_{j>i} n_j \right)
$$
of making one copy of itself,
and, finally, a probability 
$$
P_0=\frac{f}{N-1} \sum_{j>i} n_j 
$$
of making no copy at all. 

It follows that the expected number $n_i'$ of individuals in the $i$th performance class after selection is given by:
\begin{equation}
n_i'=n_i \left[ 1+ \frac{f}{N-1} \left( \sum_{j<i} n_j - \sum_{j>i} n_j \right) \right] \label{nprimo}
\end{equation}
Notice that it holds:
$$
\sum_{i=1}^M n_i'=N.
$$
Now, let us consider the mutation step. We assume that the individuals in each performance class $i$ share the same probability $Q_i$ of undergoing neutral mutations only, or no mutations at all.
We will call $Q_i$, with a slight abuse of terminology, the fidelity rate of the $i$th performance class.
Let us denote with $g_{ij}$ the probability that an individual in the $j$th performance class gives raise to an individual in the $i$th performance class as a result of a mutation ($g_{ii}=0$ since we included the neutral mutations in the fidelity rate $Q_i$. 
%\footnote{
Obviously, we could define $Q_i$ as the probability of undergoing no mutations at all and $g_{ii}$ as the probability
for the intervening mutations of being neutral. However, even if the alternative chosen in the text could seem clumsier it is better suited for our mathematical analysis.%}
).
This mutation mechanism gives raise to the following deterministic discrete equation:
\begin{equation}
n_i''=n_i' Q_i +\sum_{j=1}^M (1-Q_j) n_j' g_{ij}. \label{nsecondo}
\end{equation}
Notice that, since by definition, 
$$
\sum_{i=1}^M g_{ij}=1, \label{geq1}
$$
it will also hold
$$
\sum_{i=1}^M n_i''=N.
$$

\begin{figure}[!ht]
\ifcase\nofigure
\hspace{5mm}\includegraphics{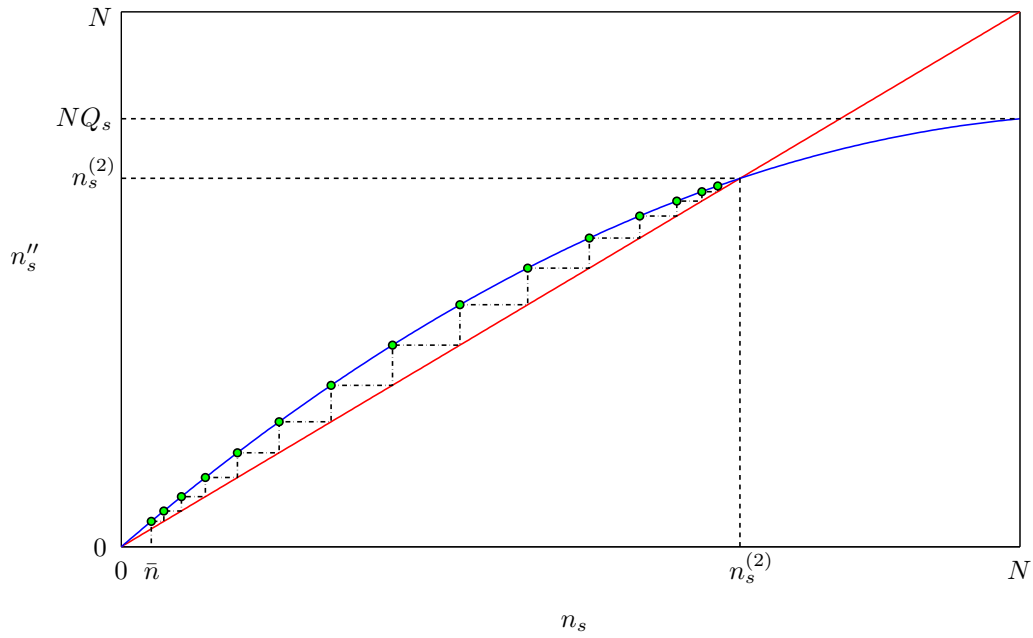}
\or
\fi
\caption{{\bf Stable state for the occupation number of the highest occupied performance class.} The blue curve represents the function $n_s''(n_s)$ defined by (\ref{nss}), while the red line corresponds to $n_s''=n_s$. The green points represent $15$ iterates of the 
discrete map (\ref{nss}) starting by the initial datum $\bar{n}$. The asymptotic value of the map (\ref{nss}) will be $n^{(2)}_s$ for any initial 
datum $\bar{n} \neq 0$.
\label{dinamica}}
\end{figure}

Suppose now that $g_{ij} \ll 1$ if $i> j$, namely that the probability of a mutation to a higher performance class is very small, then the fraction of individuals undergoing a beneficial mutation 
in one generation is negligible. 
Let $s$ be the best occupied performance class at a given time $n_s>0$, $n_i=0, \ i>s$, and suppose $1<s<M$. From equation (\ref{nprimo}), 
we get that it also holds $n_s'>0$, $n_i'=0, \ i>s$. Then, from (\ref{nsecondo}) and (\ref{nprimo}) we get:
\begin{equation}
n_s''=n_s' Q_s= n_s Q_s \left[ 1+ \frac{f}{N-1} \left( \sum_{j<s} n_j \right) \right]= n_s Q_s \left[ 1 + \frac{f (N-n_s)}{N-1} \right] \label{nss}
\end{equation}
The best performance class %$n_s''$ 
is stably populated if $n_s''=n_s$. We have two solutions. 
The first one is given by $n_s^{(1)}=0$ and the second by
\begin{equation}
n_s^{(2)}=\frac{1}{f} \left( N (1+f)-1-\frac{N-1}{Q_s} \right). \label{ns2}
\end{equation}
$n_s^{(2)}$ is greater than zero if 
$$
Q_s>\frac{1}{1+f\frac{N}{N-1}}=\frac{1}{1+f}+O \left(\frac{1}{N} \right) 
$$  
and in such a case $n_s=n_s^{(2)}$ is a sink and $n_s=0$ an unstable equilibrium, since the function $n_s''-n_s$ is positive for $n_s \in (0,n_s^{(2)})$ and negative for $n_s \in (n_s^{(2)},N)$,
as shown in figure \ref{dinamica}. 
If 
$$
Q_s<\frac{1}{1+f\frac{N}{N-1}}, 
$$
then there is only a sink in $n_s=0$. Hence, the error threshold is given by 
\be\label{qt}
\bar{Q}=1-\bar{P}=\frac{1}{1+f\frac{N}{N-1}}
\ee
Neglecting the $O(1/N)$ corrections, we obtain:
\begin{equation} \label{errort}
\bar{P}=\frac{f}{1+f}
\end{equation}
This is the same result that one gets from the Eigen model when considering the single peak fitness landscape with $A_1=(1+f)(A_2-D_2+D_1)$ 
(see equation (\ref{Pt}) and also \cite{Nowak}).

With the previous argument we have shown that, after infinitely many generations, the best occupied performance class must satisfy $Q_s>\bar{Q}$ namely that $n_j=0$ for all $j$ such that $Q_j<\bar{Q}$.
We can now show that actually the index $s$ is actually the largest possible one, namely that there is no class $i$ 
such that $Q_s>Q_i>\bar{Q}$. 
Indeed, let us suppose that $g_{i+1,i} \neq 0 \ \forall\ i=1,\dots,M-1$, then if at a given generation, the $i$th performance class is populated, while the $i+1$th is empty, then at the following generation we will have 
\begin{eqnarray}
n_{i+1}''&=&\sum_{l\leq i}(1-Q_l)\ g_{i+1,l}\ n_l\left[ 1+ \frac{f}{N-1} \left( \sum_{j<l} n_j - \sum_{j>l} n_j \right) \right]\nonumber \\
&\geq&
(1-Q_i)\ g_{i+1,i}\ n_i\left[ 1+ \frac{f}{N-1} \sum_{j<i} n_j \right]>0 \label{riempi}
\end{eqnarray}
So, a fraction of the population (possibly very small) will filtrate progressively into higher performance classes. This process will continue
until the last performance class $M$ or a performance class $s$ such that $Q_s>\bar{Q}$, $Q_i<\bar{Q}$ if $i>s$ will be reached.
Then the asymptotic occupation number of this class will be given by equation (\ref{nss}). A certain number of observations about this result are in order. First, let us notice that according to equation (\ref{riempi}), the $s+1$th performance class will be populated at each 
generation by mutants of the $s$th one. As we said, if $g_{s+1,s}$ is small, this number will be a tiny fraction of $n_s$ and we can neglect it (as we did in equation (\ref{nss})). So, what we have really shown is that the $s$th performance class is the last one that 
will have a significative occupation number. The actual value of this number will depend mainly on how near $Q_s$ is to $\bar{Q}$.

The second argument to keep into account is that the time to populate the $s$th class could be astronomical and will depend on the values of $g_{ij}$, $Q_i$ and $f$. In particular, to keep it reasonable, $g_{i+1,i}$ and $f$ must not be exceedingly small.
It is also necessary that for $i<s$ the fidelity rates $Q_i$ are not smaller than or too near to $\bar{Q}$.
A natural assumption avoiding this occurrence is that $Q_i$ is a monotonically decreasing function of $i$.

\begin{figure}[!ht]
\ifcase\nofigure
\hspace{13mm}\includegraphics{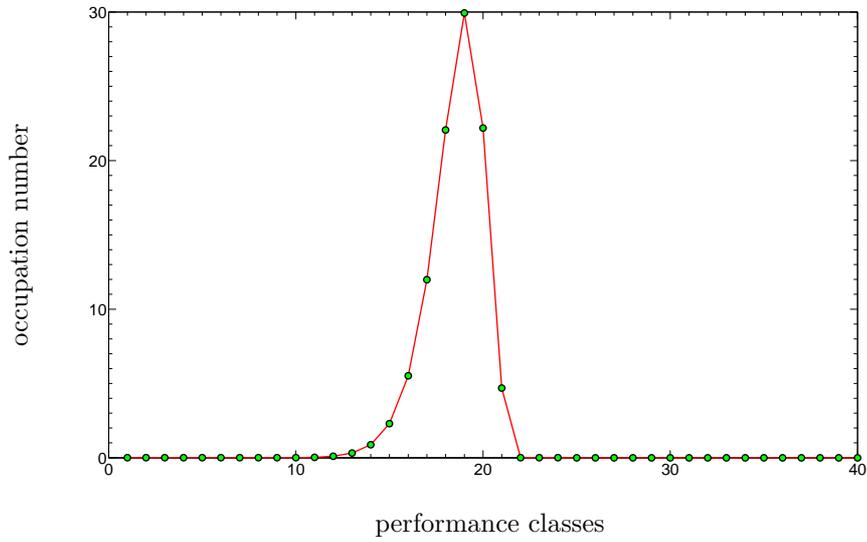}
\or
\fi
\caption{{\bf Results of the numerical simulation described in section ``The deterministic model''.} The green points show the occupation numbers of the $40$ performance classes 
obtained after $2 \cdot 10^6$ generations while the red line connect those obtained after $10^6$ generations. The prediction in (\ref{qt}, \ref{quali}) for the best occupied performance class is $s$=21.\label{simul}}
\end{figure}

As an illustrative example, we show in figure \ref{simul} the results of a numerical simulation of the discrete system (\ref{nprimo}), (\ref{nsecondo})
with the following choices of the parameters:
$M=40$, $N=100$, $f=10^{-3}$,  
$$
g_{ij}=\left\{
\begin{array}{ll}
1-10^{-6}, &\qquad \mbox{if } i=j=1 \mbox{ or } j-i=1 \\
10^{-6}, &\qquad \mbox{if } i=j=40 \mbox{ or } i-j=1 \\
0, &\qquad \mbox{otherwise},
\end{array} 
\right.
$$ 
\be 
Q_i=(1-10^{-5})^{\sqrt{i^3}}, \qquad i=1,\dots,40. \label{quali}
\ee
and with all the $100$ individuals in the first performance class as an initial state. Needless to say that this choice of the parameters does not pretend to have any degree of
biological realism. 
The green points show the occupation numbers of the $40$ performance classes obtained after $2 \cdot 10^6$ generations, while the red line connect those obtained after $10^6$ generations. The maximum 
difference between the occupation numbers of the same performance class at $2 \cdot 10^6$ and $10^6$ generations is $5 \cdot 10^{-5}$ and, consequently, cannot be detected.   
This means that the population had almost reached its stable state after $10^6$ generations.   
The error catastrophe does occur when the fidelity rate (\ref{quali}) is less than the fidelity threshold (\ref{qt}).
With the above choices, we have $Q_i>\bar{Q}$ for $i=1,\dots,21$ and $Q_i<\bar{Q}$ for $i=22,\dots,40$. According to the above,
we expect that, in the asymptotic limit, the performance classes from the $22$nd on, should be empty, while equation (\ref{ns2}) predicts that the $21$st one should be occupied by $4.682$ individuals. At the end of the simulation,
the number of individuals in the $21$st performance class  is $4.686$ while those in the $22$nd are $2.02 \cdot 10^{-4}$ and they go progressively decreasing, by approximately $5$ orders of magnitude per performance class, while the performance class increases.

\begin{figure}[!ht]
\ifcase\nofigure
\hspace{30mm}
\includegraphics[scale=0.6]{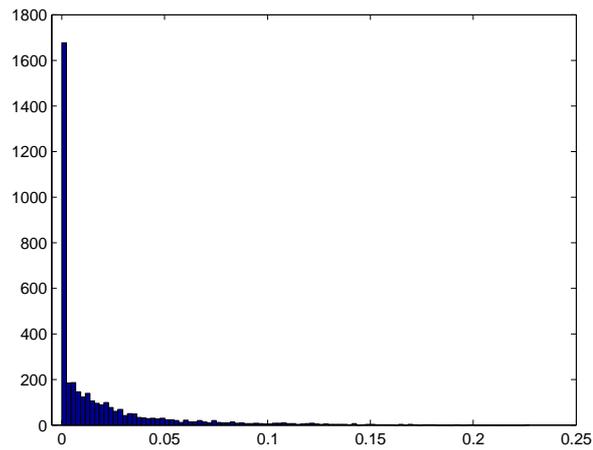} 
\or
\fi
\caption{{\bf Histogram of the distribution of $\sigma/\bar{\nc}$ for the $3740$ runs of our simulations.} $\bar{\nc}$ is the average number of coding triplets for the best individuals in the last generation
and $\sigma$ is the corresponding standard deviation. The size of the bins is $0.0025$.
\label{sigmaNc} }
\end{figure}

\paragraph{The TMs critical number of coding states}
We made two hypotheses in our deterministic model to find the value of the error threshold (\ref{errort}). The first hypothesis is that $g_{ij} \ll 1$ if $i > j$, that is, that favorable mutations are extremely rare. This is a natural 
assumption in our TMs model, because very often the mutations induce a big change in the output tape that have a very small probability of being favorable. Moreover, in the next section, we will develop a stochastic model based on the same assumption, and we will see (figure (\ref{ext1})) that there is good agreement between the prediction of this model and the observed results. 
The second relevant assumption to compute the error threshold (\ref{errort}) is that the individuals belonging to the best performance class $s$ have the same 
fidelity rate $Q_s$. 
We will make the further assumption that, for TMs, the probability that a mutation occurring in a coding triplet is neutral is also negligible. Then, the fidelity rate of a 
TM with $N_c$ coding triplets is given by  
\begin{equation}
Q=\left( 1- \pmm \right)^{3 N_c}, \label{qualityTM}
\end{equation}
since mutations occurring in non-coding triplets are, by definition, neutral. 
It follows that, for a given value of $\pmm$, the fidelity rate is determined only by the number $\nc$ of coding triplets of the TM. The assumption that 
the best performing TMs have the same fidelity rate is therefore equivalent to the assumption that they have the same number of coding triplets. 
Figure \ref{sigmaNc} shows that this assumption is very near to the truth when considered for a given run (that is what we really need). 
However, notice that the relation among $s$ and $Q_s$ varies considerably  among different runs. 
This is particularly evident in figure \ref{correlation}, where the number of coding triplets associated with the performance scores of $47$ and $48$ exhibits a more than two-fold variation. 

Having established that the hypotheses under which we have obtained the error threshold (\ref{errort}) are accurate for our model, we can use it to determine the 
maximum allowed number of coding triplets for a TM.
In the case of our  model, $f=1$, so that equation (\ref{errort}) give us the error threshold at $\bar{P}=1/2$. The mutation probability for a TM with $\nc$ coding triplets is given by:
\begin{equation}
P=1-(1-\pmm)^{3 \nc}. \label{mutP}
\end{equation}
By equating (\ref{mutP}) to the error threshold, we get the critical number of coding states for the TMs:
\begin{equation}
\nc^{\ast}=-\frac{\ln(2)}{3 \ln(1-\pmm)} \label{Ncrit}
\end{equation}
This expression is represented by the thick black line in figure~\ref{nclimit}.
The ultimate fate of TMs with a number of coding states larger than $\nc^{\ast}$, according to our deterministic model, will be the extinction. 

\subsection{The stochastic model} \label{stocha}

In this section we will keep into account the stochastic effects in our mutation and selection procedures. 

We recall that  the constant population size $N$ must be an even number. 
We will introduce a stochastic model for the evolution of only the number $n_s$ of individuals with the best performance value.
Let us consider separately the selection and mutation steps.

\paragraph{The selection step}
Since we are interested in the evolution
of the number $n_s$ of the best individuals only, we can put all the remaining $N-n_s$ individuals into the same class. Let us denote with the symbol ``$1$''  
the individuals of the best performance class and with the symbol ``$0$'' all the others. We will denote by $n_s'$ the number of individuals in the highest 
performance class in the new population. $n_s'$ will be determined by the number of pairs $11$, $10$ and $00$ that we will get extracting random 
pairs without replacement from the old population. Let us denote by $k$ the number of $11$ pairs, by $l$ the number of $10$ pairs and by
$m$ the number of $00$ pairs. As a consequence we will have
\be
n_s=2k+l \qquad n_s'=2(k+l) \qquad 2(k+l+m)=N
\ee
The probability that we get $n_s'=2(k+l)$ individuals into the best class when applying the selection step to a population with $n_s=2k+l$ individuals
into the best class, is given by the probability that we extract $k$ $11$ pairs, $l$ $10$ pairs and $m$ $00$ pairs from a set containing 
$2k+l$ ones and $l+2m$ zeroes. This probability is given by:
$$
P(\overbrace{(11) \dots (11)}^k \overbrace{(10) \dots (10)}^{l} \overbrace{(00) \dots (00)}^{m} )= 2^l 
\left( 
\begin{array}{c}
k+l+m\\
k
\end{array}
\right)
\left( 
\begin{array}{c}
l+m\\
l
\end{array}
\right)
\left/ 
\left( 
\begin{array}{c}
2(k+l+m)\\
2k+l
\end{array}
\right)
\right.
$$
Indeed, the $2^l$ term keeps into account that the $l$ pairs $10$ can be obtained extracting the $1$ before the $0$ or vice versa. 
The term 
$$
\left( 
\begin{array}{c}
k+l+m\\
k
\end{array}
\right)
$$
gives the number of possible distributions of the $k$ $11$ pairs inside the $k+l+m$ total pairs.
The term 
$$
\left( 
\begin{array}{c}
l+m\\
l
\end{array}
\right)
$$
gives the number of possible distributions of the $l$ $10$ pairs inside the remaining $l+m$ pairs.
Finally,
$$
\left( 
\begin{array}{c}
2(k+l+m)\\
2k+l
\end{array}
\right)
$$ 
is the number of possible distributions of the $2k+l$ $1$ symbols in the $2(k+l+m)=N$ possible places.

Let us notice that   
\begin{itemize}
\item $n_s'$ is always even, 
\item $n_s' \geq n_s$,
\item $n_s' \leq 2 n_s$.
\end{itemize}  
If we fix $n_s$ and $n_s'$ satisfying the above constraints, we can obtain $k$ and $l$ as a function of $n_s$ and $n_s'$:
$$
k=\frac{2 n_s-n_s'}2 \qquad l=n_s'-n_s
$$
Since the total population is fixed to $N$ we have:
$$
2(k+l+m)=N \quad \Longrightarrow \quad m=\frac{N-n_s'}{2}
$$
Hence, the probability of getting $n_s'$ individuals into the best performance class after applying the selection procedure to a population
with $n_s$ individuals into the best performance class is:
$$
P_{rip}(n_s \to n_s')= \left\{ 
\begin{array}{l}
\displaystyle 0 \qquad {\rm if} \ n_s'<n_s \ {\rm or} \  n_s'>{\rm min}(2 n_s,N) \\
\displaystyle 0 \qquad {\rm if} \ n_s'\ {\rm odd}\\
\displaystyle 2^{(n_s'-n_s)} \left( 
\begin{array}{c}
\frac{N}2\\
\frac{2n_s-n_s'}2
\end{array}
\right)
\left( 
\begin{array}{c}
\frac{N-2 n_s+n_s'}2\\
n_s'-n_s
\end{array}
\right)
\left/ 
\left( 
\begin{array}{c}
N\\
n_s
\end{array}
\right)
\right.
 \qquad {\rm otherwise}.
\end{array}
\right.
$$

\paragraph{The mutation step} 
Let us introduce mutation into the model. We will follow to use the two simplifying assumptions that we used for the deterministic model, namely:
\begin{enumerate}
\item TMs in the best performance class have the same number of coding triplets $\nc$. 
\item Mutations in coding triplets are (almost) always deleterious. 
\end{enumerate} 
By definition, mutations in non-coding triplets are neutral.

Under this assumptions if $n_s'$ is the number of best individuals before mutation, the probability of getting $n_s''=n_s'-k$ individuals after the mutation 
step is given by
$$
P_{mut}(n_s' \to n_s''=n_s'-k)=\left(
\begin{array}{c}
n_s' \\
k
\end{array}
\right) P^k (1-P)^{n'_s-k},
$$ 
where we denoted by $P$ the probability that an individual in the best performance class will undergo at least one mutation into a coding triplet. 
$$
P=1-(1-\pmm)^{3\nc}
$$
\paragraph{The Markov matrix} 
If the total population $N$ is finite, under our assumption the best individuals will always go extinct in a finite time. 
The expected number of generations $\tau$ before it happens can be computed using  the Markov matrix $M$ of the process \cite{AMS}. 
The entries $M_{ij}$ of the Markov matrix give the probability that the system under scrutiny pass from its $i$th state to the $j$th one.
In our case the state of the system is labeled by the number $n_s$ of individuals into the best performance class and the entries of $M$ will be given by
$$
M_{n_s+1,n_s''+1}=\sum_{n_s'=0}^N P_{rip}(n_s \to n_s') P_{mut}(n_s' \to n_s''), \qquad n_s,n_s''=0,\dots,N 
$$
The state $n_s''=0$ will be an absorbing state for $M$ and the procedure to compute the expected number of generations $\tau$ for reaching it, works as follows. Let $S$ be the matrix that one obtains by removing the first row and the first column corresponding to the only absorbing state and let 
$\bf{c}$ be a $N-$dimensional vector whose entries are all one. The matrix $\mathbb{I}-S$, where $\mathbb{I}$ denotes the identity matrix, is invertible. If the Markov process begins in the state $i$,
then the expected number of generations before extinction will be given by:
\be
\tau=\left[(\mathbb{I}-S)^{-1} {\bf{c}}\right]_i \, . \label{eqt}
\ee
Let us stress that the equation (\ref{eqt}) is obtained by assuming that the system evolves for an infinite number of generations.
The expected extinction times versus the number of coding triplets are plotted in figure \ref{ext1} for $5$ different values of the mutation probability.

\subsection{Simulations settings}

In this subsection we introduce the parameter values that we adopted in our computer simulations. 

We chose the goal tape containing the binary expression of the decimal part of $\pi$ (the dots 
are just a useful separator):
$$
\begin{array}{l}
0010010000.1111110110.1010100010.0010000101.1010001100.0010001101.0011000100.1100011001.\\
1000101000.1011100000.0011011100.0001110011.0100010010.1001000000.1001001110.0000100010.\\
0010100110.0111110011.0001110100.0000001000.0010111011.1110101001.1000111011.0001001110.\\
0110110010.0010010100.0101001010.0000100001.1110011000.1110001101.
\end{array}
$$ 
As a consequence, the maximum possible performance value is $125$.
We performed simulations with the following (approximate) values of the states-increase rate $\pii$ and point mutation probability $\pmm$:
\begin{eqnarray*}
\pii &\in& \left\{9.26 \cdot 10^{-5}\,; 1.66 \cdot 10^{-4}\,;3.00 \cdot 10^{-4}\,;5.40 \cdot 10^{-4}\,;9.72 \cdot 10^{-4}\,;1.75 \cdot 10^{-3}\,; \right.\\
&& \left. \ 3.14 \cdot 10^{-3}\,;5.68 \cdot 10^{-3}\,;1.02 \cdot 10^{-2}\,;1.85 \cdot 10^{-2}\,;3.33 \cdot 10^{-2}; 6.00 \cdot 10^{-2}\,; \right.\\
&& \left. \ 1.08 \cdot 10^{-1}\,; 1.95 \cdot 10^{-1}\,;
3.51 \cdot 10^{-1}\,;6.33 \cdot 10^{-1}\,;1.14\right\}\,. \\
\pmm &\in& \left\{ 4.91 \cdot 10^{-5}\,; 8.10 \cdot 10^{-5}\,; 1.34 \cdot 10^{-4}\,; 2.21 \cdot 10^{-4}\,; 3.64 \cdot 10^{-4}\,; 6.01 \cdot 10^{-4}\,; \right. \\
&& \left.  
9.91 \cdot 10^{-4}\,; 1.64 \cdot 10^{-3}\,; 2.70 \cdot 10^{-3}\,;4.44 \cdot 10^{-3}\,; 7.35 \cdot 10^{-3}\, \right\}\, .
\end{eqnarray*}   
These values have been chosen in such a way that consecutive ones have a constant ratio.
For any pair of values $\pii, \pmm$, we performed $20$ simulations varying the initial seed of the C native random number generator, for a total of $3740$ runs. Each simulation lasted $50000$ generations.

\section{Results} \label{results}

\subsection{Performance, coding triplets and mutation probabilities}

In this subsection we analyze how the performance and the number of coding triplets of the best performing machines vary with the different values of the mutation and states-increase rates.

\begin{figure}[!ht]
\ifcase\nofigure
\hspace*{12mm}\includegraphics{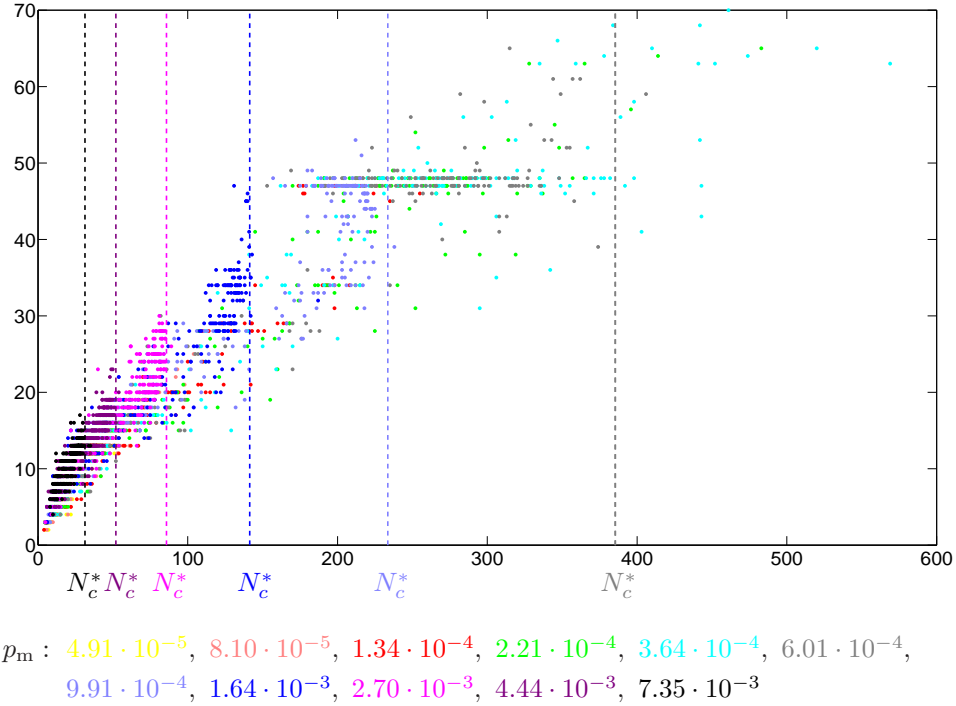}
\or
\fi
\caption{{\bf Performance versus the number of coding triplets.} The performance is shown for the best performing TMs at generation $50000$ for the $3740$ runs of our simulations. Each color corresponds to a different value of the 
mutation probability as indicated in the scale under the image. 
%$\pmm$: $\textcolor[rgb]{1,1,0}{4.91 \cdot 10^{-5}}$, $\textcolor[rgb]{1,0.5,0.5}{8.10 \cdot 10^{-5}}$, $\textcolor[rgb]{1,0,0}{1.34 \cdot 10^{-4}}$, $\textcolor[rgb]{0,1,0}{2.21 \cdot 10^{-4}}$, $\textcolor[rgb]{0,1,1}{3.64 \cdot 10^{-4}}$, $\textcolor[rgb]{0.5,0.5,0.5}{6.01 \cdot 10^{-4}}$,
%$\textcolor[rgb]{0.5,0.5,1}{9.91 \cdot 10^{-4}}$, $\textcolor[rgb]{0,0,1}{1.64 \cdot 10^{-3}}$, $\textcolor[rgb]{1,0,1}{2.70 \cdot 10^{-3}}$, $\textcolor[rgb]{0.5,0,0.5}{4.44 \cdot 10^{-3}}$, 
%$\textcolor[rgb]{0,0,0}{7.35 \cdot 10^{-3}}$. 
The dashed lines correspond to the critical number of coding triplets for the $6$ highest value of the mutation probabilities. For lower values 
the corresponding critical number of coding triplets lies outside of the graph. Notice that many points are superimposed. 
\label{correlation} }
\end{figure}

\begin{figure}[!ht]
\ifcase\nofigure
\hspace{32mm}\includegraphics[scale=0.5]{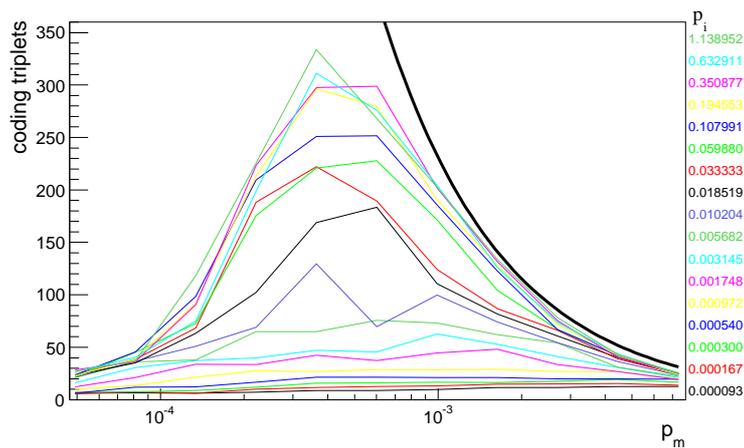}
\or
\fi
\caption{\label{nclimit} {\bf  Plot of the number of coding triplets for the best machine in the population.} 
The number of coding triplets after $50000$ generations, averaged on the seeds, is shown as a function of $\pmm$, for all the 
values of $\pii$. The black thick line on the right represents the critical 
number of coding triplets, according to equation (\ref{Ncrit}).}
\end{figure}

\begin{figure}[!ht]
\ifcase\nofigure
\includegraphics{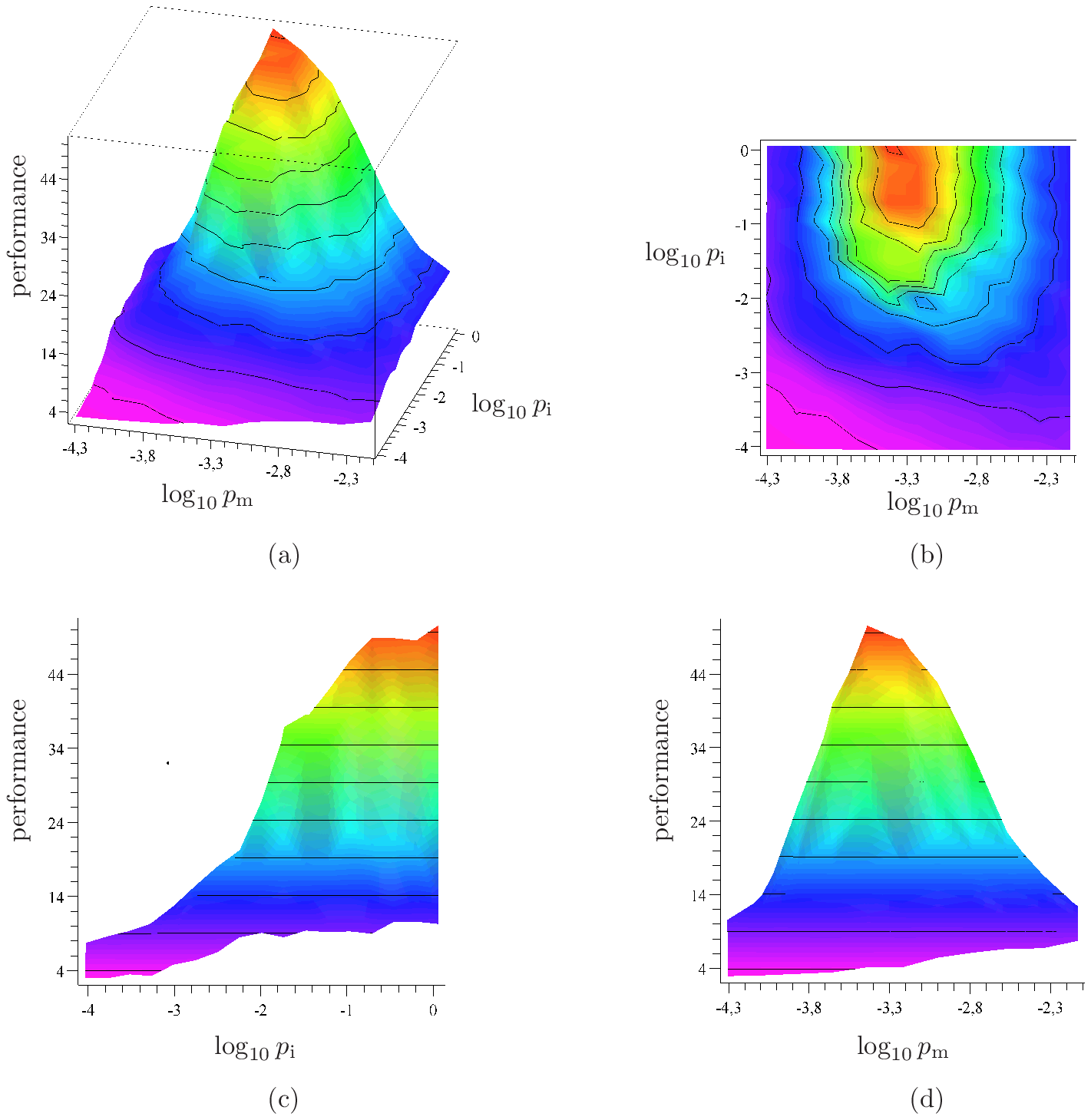}
\or
\fi
\caption{\label{performance_graph}{\bf Best performance value in the population at the last generation.} 
The best performance is averaged on the twenty different seeds and plotted as a function of the states-increase rate $\pii$ and 
of the mutation rate $\pmm$. (a) shows a 3D view, while subfigures (b), (c), (d) correspond to the three orthogonal projections.}
\end{figure}

\begin{figure}[!ht]
\ifcase\nofigure
\hspace{40mm}\includegraphics{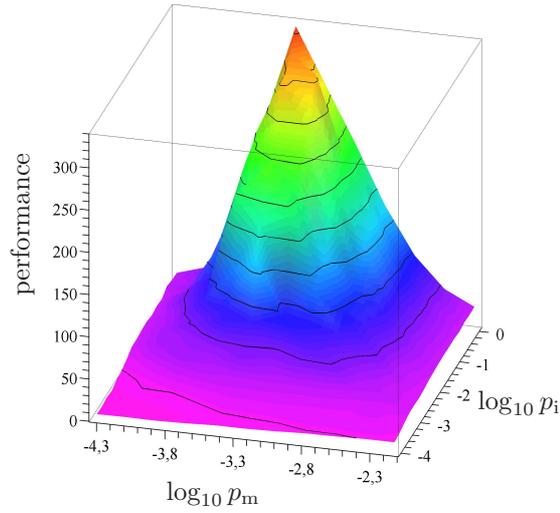}
\or
\fi
\caption{\label{esoni} {\bf Number of coding triplets in the population at the last generation.}
The number of coding triplets is averaged on the best machines and on the seeds; it is plotted versus $\pii$ (right) and $\pmm$ (left).}
\end{figure}

In figure \ref{performance_graph} we plot the best performance value obtained in the population at the last generation (averaged on the different choices of the seed) versus the state-increase rate $\pii$ and the mutation probability $\pmm$. The maximum performance value of $50.6$ is obtained for the maximum value of $\pii$, $\pii \simeq 1.14$ (see figure \ref{performance_graph}.c) and an intermediate value of $\pmm$, $\pmm \simeq 3.64 \cdot 10^{-4}$ (see figure \ref{performance_graph}.d). In figure \ref{esoni}, we show the number of coding triplets $\bnc$ (averaged on the best performing machines at the last generation and on the seeds), versus $\pii$ and $\pmm$. Again the maximum value $\bnc=333.9$ is obtained for exactly the same values of $\pii \simeq 1.14$ and $\pmm \simeq 3.64 \cdot 10^{-4}$. This fact and the similarity between figure \ref{performance_graph}.a and \ref{esoni} suggest a strong correlation between the performance and the number of coding triplets. Indeed, the correlation coefficient between them is $r=0.95$ (see also figures \ref{correlation}, \ref{doppio_grafico}, \ref{correlazione}). The fact that the maximum performance occurs
for an intermediate value of $\pmm$ is also partially due to this correlation. Indeed, if the mutation probability is too low, there is no enough variability among the TMs 
for selection to work on, while when the  mutation probability is too high, the error threshold exerts a strong limiting action on the maximum number of coding triplets.
This latter effect is clearly visible in figures \ref{correlation} and \ref{nclimit}, both taken after 50000 generations. 
Indeed, in figure \ref{correlation}, the abscissa positions seldom exceed the corresponding vertical lines at $\nc^{\ast}$.  
The presence of the error threshold also affects the trend of the performance with the generations. Indeed, when both
$\pmm$ and $\pii$ are large, the TMs approach very early the maximum number of coding triplets. From that moment on, further accumulation of coding triplets is strongly 
opposed by mutation and selection. This leads to a saturation in the performance and in the number of coding triplets that is clearly visible in the plateau of figure \ref{doppio_grafico}, (b) and (d). Notice that this plateau effect is not present when  
$\pii$ is small (figure \ref{doppio_grafico}, (a) and (c)). 
%This behaviour suggests that the optimal choice for the mutation probability should be adaptative, varying with the number of coding triplets during the generations. 
This behaviour suggests that an adaptative choice for the mutation probability could maximize the speed of evolution. 
Indeed, one could start with an high mutation rate in the first generations to increase the variability, progressively diminishing it when the number of coding triplets increase to reduce the limiting effect due to the error threshold.
We presented a proposal for the optimal adaptative mutation probability for this model in \cite{Ideal}.  

Figure \ref{correlazione} shows, on a log-log scale, the relation between $\bnc$ and $\pii$. The straight line of linear regression has been evaluated in the range $\pii \leq 3.33\cdot 10^{-2}.$ The reason is that this range corresponds to the one considered in \cite{PRE}, allowing us to compare the two results. Moreover, it is clear that the linear regime does not hold for large values of $\pii$, for which we
observe a saturation effect. The regression gives the relation 
\begin{eqnarray}
\bnc=7.3 \cdot 10^2\ \pii^{0.46}, && \mbox{present simulations,}\nonumber \\
\bnc=2.5 \cdot 10^3\ \pii^{0.53}, && \mbox{paper \cite{PRE},}\label{scala}
\end{eqnarray}
both exponents being close to $\frac12$. Now, if $\bnt$ is the total number of states, its expected value is %\footnote{This is true in absence of selection but, as discussed in \cite{PRE}, it remains approximately true even with selection, except for very small values of $\pii$.}
\be
\bnt=50000\cdot \pii+1 \quad \Rightarrow \quad \pii \simeq \frac{\bnt}{50000},
\ee
(this is true in absence of selection but, as discussed in \cite{PRE}, it remains approximately true even with selection, except for very small values of $\pii$)
so that 
\be
\frac{\bnc}{\bnt} \propto \frac1{\sqrt{\pii}}
\ee
%Putting this in (\ref{scala}) we get that the number of coding triplets
%scales almost as the square root of the total number of states
%\be
%\bar{N}_c\propto \bar{N}_t^{0.46}
%\ee
This means that the fraction of coding triplets on the total will decrease when $\pii$ increases (strictly speaking, this analysis holds only for the linear regime however it is clear that, 
for larger values of $\pii$, the plateau of figure \ref{correlazione} corresponds to an amplification of this effect).  

As in our previous paper \cite{PRE}, we observe that the maximum performance is obtained for the maximum value of $\pii$. However, this time, the trend of the performance 
with $\pii$ is not strictly monotonically increasing, since, as shown in figure \ref{performance_graph}.c, there is a plateau for high values of $\pii$. 
Notice that this plateau corresponds exactly to the region in figure \ref{correlazione} where the number of coding triplets reaches a saturation, so that this effect also is due to the 
presence of the error threshold. 
For various reasons, explained in \cite{PRE}, we believe that
the performance should exhibit a maximum for a finite value of $\pii$ (this is one of the reasons that led us to increase upward the range of variation of $\pii$). Unfortunately,
it seems that if this maximum exists, it lies outside of the range of values of $\pii$ that we selected. Finally, it is interesting to compare figure \ref{performance_graph}.c 
restricted to the range of $\pii$ values considered in \cite{PRE} with the figure 3.c of \cite{PRE}, that we reproduce here (see figure \ref{comparison}).
Despite the fact that we changed the number of generations (they were $200000$ in \cite{PRE}), the way the head can move on the tape (in \cite{PRE} it could also stay still) and the topology 
of the tape (in \cite{PRE} it was not periodic), the 
two profiles are very similar. This means that the dependence of the performance on $\pii$ is very robust in this model. That's not the case of the dependence
of the performance on $\pmm$, that is influenced, for example, by the choice of the number of generations. 

The trend of the performance with $\pii$ is interesting because it 
suggests that, in our model, the presence of inactive and free to mutate code strongly improves the evolvability of our populations.

\subsection{Extinction times}

In figure \ref{ext1}, we plot the base $10$ logarithm of $\tau$, the expected number of generations before extinction given by equation (\ref{eqt}) versus the number of coding triplets (red line), and superimpose the data
obtained from our simulations (blue points). The red line is obtained numerically  
starting with a unique individual in the best performance class ($i=1$ in (\ref{eqt})) through the Markov matrix of the process, as explained in the previous section. 
The blue points give the observed value of $\tau$ averaged over bins through the following procedure. First, 
the whole range of the number of coding triplets is divided into bins.
The size of the bin is different for the different values of the mutation probability 
$\pmm$ and is given by the smallest integer greater than or equal to the critical value for the number of coding triplets (see eq. (\ref{Ncrit})) divided by $40$. It is necessary to introduce bins because otherwise, especially when $\nc^\ast$ is
large, there are too few extinction events associated with any value of the coding states to give raise to an also minimal statistics. On the other hand, one has to avoid that the bin size is so large
that the expected number of generations before extinction varies considerably inside it. It seems that dividing $\nc^\ast$ by $40$ is a good choice for the bin size. 

Now, let us suppose that at a given generation $\tau'$, the data register a drop in the maximum performance value, then we go back to the generation $\tau$ when this performance value appeared and
we count the number of TMs scored with it. If this number is exactly two, then we register the extinction time $\tau'-\tau$ and increment the number of extinction events registered in the bin containing the
number of coding states of the two TMs at the generation $\tau$.         
The discrepancy between the fact that we compare the extinction data obtained starting with two individuals
in the best performance class with those obtained from the Markov process starting with a unique individual is due to the fact that our program registers the data after the selection step, when the best performing individual has already made a copy of itself.
%\footnote{
We are neglecting the quite improbable case that after mutation two new best performing individuals (with the same performance) do emerge and they are extracted as a pair in the subsequent selection step. %}.
If the number of extinction events registered for a bin is greater than or equal to $5$, then a blue point is plotted with an $x$ coordinate equal to the center of the bin and 
an $y$ coordinate equal to the mean of all the registered times of extinction. The requirement to have at least $5$ extinction events in each bin is due to the fact that 
the expected number of generations before extinction corresponds to the mean over an infinite number of extinction events. This pushes us to select a minimum number of extinction events 
as large as possible, to reduce the stochastic noise. On the other hand, if we choose a too large number, then we get too few points from our data. Again, to fix the minimum number of extinction events per bin to $5$ seemed to us a reasonable compromise.     

The figure \ref{ext1} corresponds to the $5$ largest values of the mutation probability $\pmm$ considered in the simulations. For smaller values 
of $\pmm$, too few ``experimental'' points are obtained. We see that the agreement between the theoretical model and the simulation data is extremely good from large values of $\nc$ to the peak of the blue points, that occurs when the expected number of generations before extinction is near $100$. On the left of such peak, the agreement is 
completely lost and a peculiar monotonic growth of $\tau$ appears instead. The main reason is that we run our simulations
for $50000$ generations, while the theoretical model assumes an infinite number of them. This implies that the agreement between the data and the theoretical model will be good until 
when $50000$ generations is a good approximation to $\infty$, that is when $50000$ generations is much larger than the expected number of generations before extinction. 
Clearly, when this latter number increases, the approximation is doomed to worsen. Indeed, when the theoretical model predicts that the expected number
of generations before extinction is larger than $50000$ (around $y=4.7$ in our graphs), the agreement between the model and the simulations is impossible.

We suggest two possible mechanisms 
to explain why in the region where there is no agreement with the theoretical model, the observed extinction times increase while increasing the number of coding triplets. 
The first mechanism is that, in this region, there is a relatively high probability of extinction when the best TMs have just emerged. Indeed, we know that at the beginning there will be
only two TMs in the best performance class (otherwise we do not register the data, as we explained above). If both TMs undergo a mutation (in a coding triplet) in the next mutation phase, then
they will most probably go extinct. On the other hand, if they start to spread into the population, then extinction becomes more and more improbable and, on consequence, the time to wait to observe it largely increases. In these cases, the cut to $50000$ generations will
throw away a considerable portion of the extinction probability distribution, with the effect of amplifying the weight of the 
probabilities before that generation. This effect is clearly visible in figure \ref{estinzione} where we plotted for $\pmm=4.44 \cdot 10^{-3}$ and five different values of $\nc$, the extinction probability distribution renormalized to $1$ in the range
between $1$ and $32768$ generations (this number is dictated by computational reasons). We see that the relative probability of observing an extinction event in the first few generations
decreases with $\nc$ below $\nc^*$ and increases after, in accordance with figure \ref{ext1}. 

Another mechanism can amplify this effect. 
Indeed, if the TMs extinction time is large, the probability that an increase in the performance value will occur does also increase. In such a case, the extinction of the original TMs simply will not be registered, creating a bias toward short extinction times.

\subsection{The route to the error threshold through punctuated equilibria}

Let us notice that all the possible output tapes (and consequently all the possible performance scores) can be obtained with a $300$ states TM with $300$ coding triplets and running for 
$300$ time steps. Indeed, let us denote 
with $o$ the desired output tape and with $o_i$ the entry of its $i$th cell, then the following $300$ states TM will produce it:
$$
\begin{array}{|c|c|}
\hline
 &\bf{i}  \\ \hline
 0 & o_i -  \mbox{Right}-\mbox{\bf i+1} \\ \hline
 1 & *   - \ \ \ * \  \ \   -  \ \,  {\bf * } \ \, \\ \hline
\end{array}
\quad {\bf i}=1,\dots,299, 
\qquad 
\begin{array}{|c|c|}
\hline
&\bf{300}  \\ \hline
 0 & o_{300} - * -\mbox{\bf Halt} \\ \hline
 1 & * \ \,  - *  -  \ \,  {\bf * } \ \, \\ \hline
\end{array} 
$$
Here the $*$ symbol means that the corresponding entry of the state is irrelevant. Notice also that this is only a possible solution
and, quite probably, not the shortest one.  
 
\begin{figure}[hb]
\ifcase\nofigure
\includegraphics{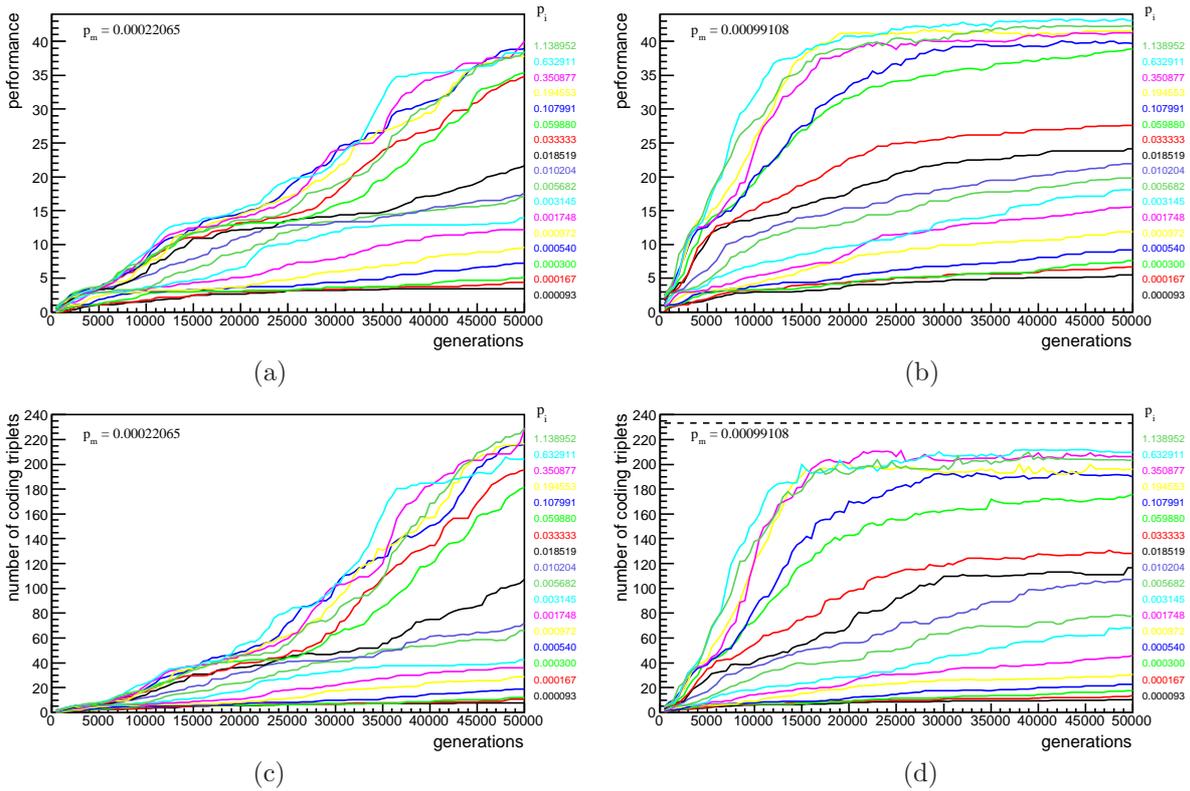}
\or
\fi
\caption{\label{doppio_grafico}{\bf Data along the generations.} 
Here we show the evolution of the performance (top) and of 
the number of coding triplets (bottom) with the generations, for 
the values of $\pii$ indicated by the matching colours and for two values of $\pmm$.
In (d), the dashed line represents the maximal number of coding triplets (\ref{Ncrit}). In (c), 
the corresponding line is outside the graph, being $\nc^{\ast}=1047.0$.
Data are sampled every 100 generations and averaged on the seeds.}
\end{figure}

\begin{figure}[!ht]
\ifcase\nofigure
\hspace{35mm}\includegraphics[scale=0.5]{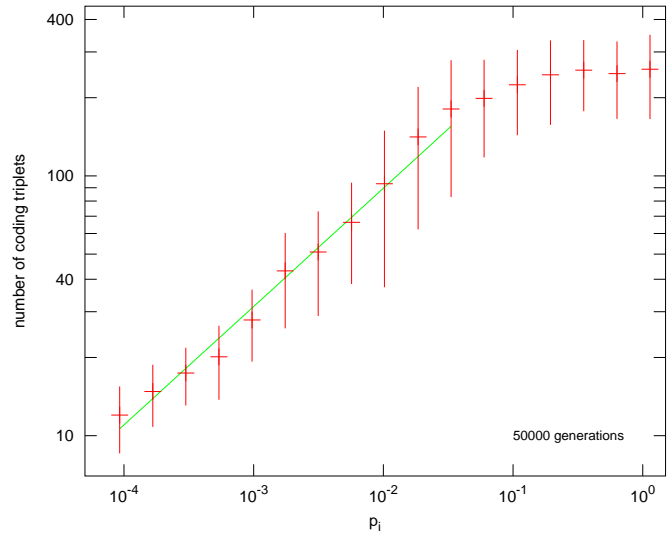}
\or
\fi
\caption{{\bf Correlation of the mean number of coding triplets $\bar{N}_c$ versus the states-increase probability $\pii$.}
For each value of $\pii$, only the four best values of the final performance (at four different $\pmm$) are retained for the evaluation of $\bar{N}_c$, for each seed. The green straight line of linear regression is evaluated on the range $\pii \leq 3.33 \cdot 10^-2$ only, in order to compare with figure 7 of \cite{PRE}.\label{correlazione}}
\end{figure}

\begin{figure}[!ht]
\ifcase\nofigure
\includegraphics{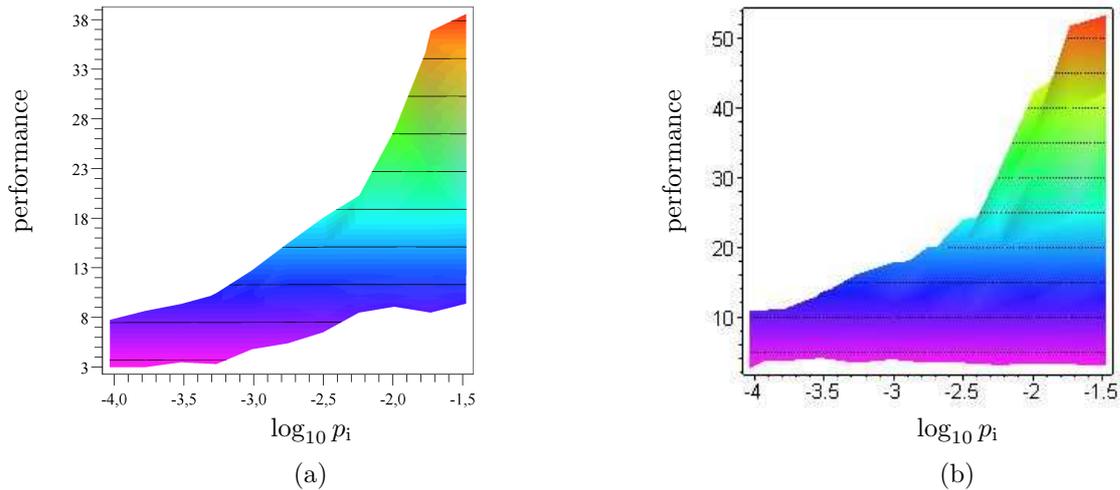}
\or
\fi
\caption{{\bf Comparison between actual and previous data.} The subfigure (a) corresponds to figure 4.c restricted to the range of $\pii$ values considered in \cite{PRE}. Subfigure (b) is the same as (a) but for the data obtained
in \cite{PRE}. \label{comparison}}
\end{figure}

\begin{figure}[!ht]
\ifcase\nofigure
\hspace{10mm}\includegraphics{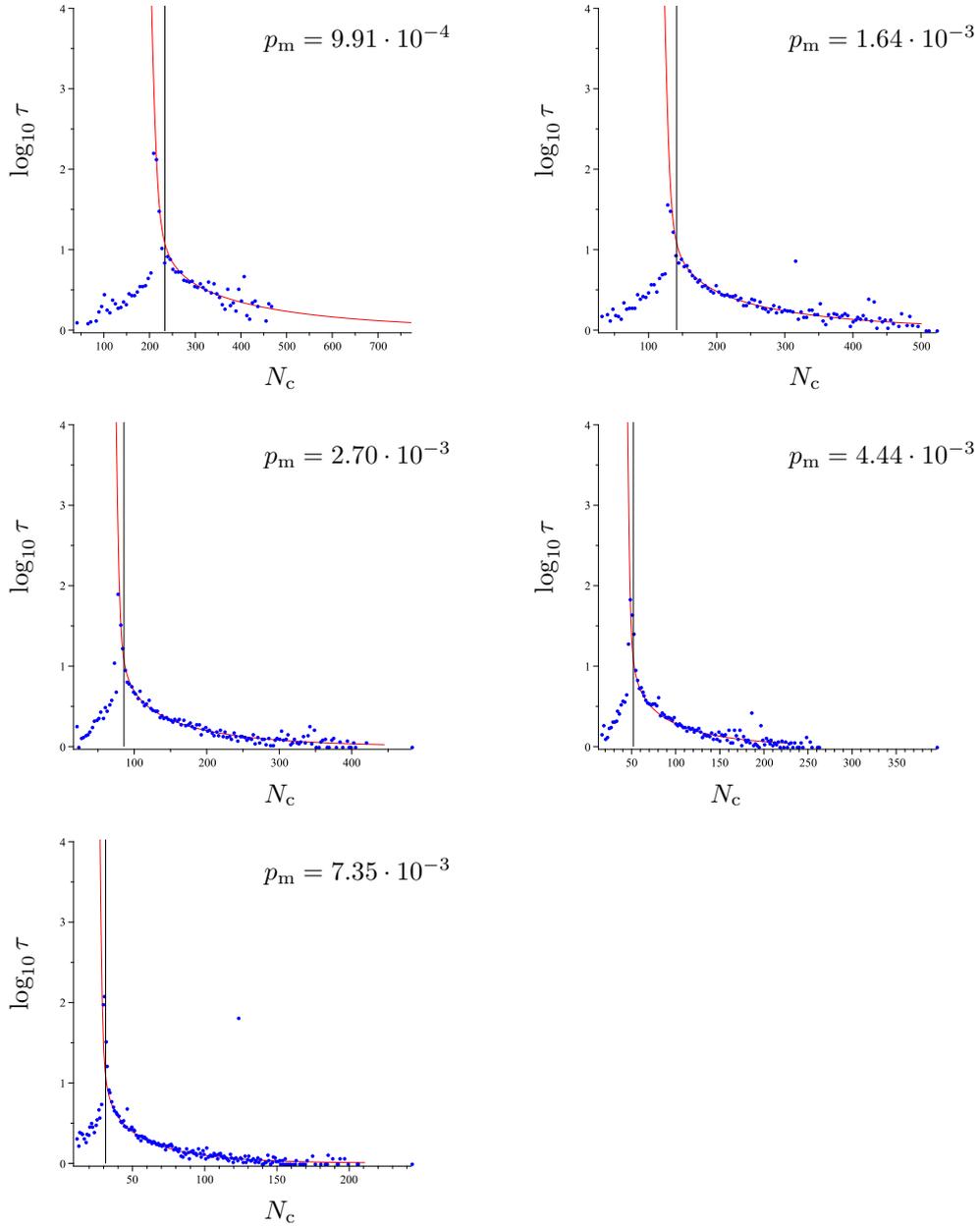}
\or
\fi
\caption{{\bf Extinction times vs mutation probabilities.} Logarithm of the theoretical (continuous line) and observed (points) extinction time $\log_{10} \tau$ versus the number of coding triplets $\nc$ for the indicated mutation probabilities. 
The black vertical lines correspond to $\nc^{\ast}$, the critical number of coding triplets of the deterministic model (eq. (\ref{Ncrit})).
\label{ext1}}
\end{figure}

\begin{figure}[!ht]
\ifcase\nofigure
\hspace{10mm}\includegraphics[scale=1]{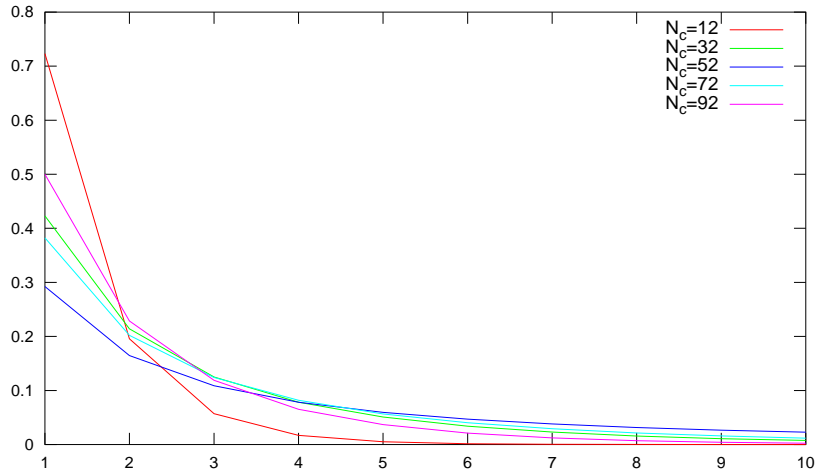}
\or
\fi
\caption{{\bf Relative extinction probabilities versus the generation number.}
The curves correspond to $\pmm=4.44 \cdot 10^{-3}$, for five different values of $\nc$. The error threshold for the given $\pmm$ is 
$\nc^{\ast}\sim 52$, represented by the blue line.
Data are renormalized by setting to one the probability of observing an extinction event before generation $32768$.\label{estinzione}} 
\end{figure}

Figure \ref{correlation} shows that, for some of 
the values of the mutation probabilities, TMs can indeed accumulate $300$ coding triplets, or even more, during the $50000$ generations. So, the TMs could, in principle, attain the maximum possible performance value of $125$. However, the 
actual maximum performance value obtained in the $3740$ runs of our simulations is $70$, quite far from the theoretical maximum.
This is due to the fact that TMs do not optimize the use of coding triplets as it is apparent from figure \ref{correlation}. Indeed, if we 
consider the various TMs that obtain a performance value of $47$ (for example), we see that the number of coding triplets spans a range
from $131$ to $443$. It is worth noticing that by diminishing the mutation probability, the number of coding triplets tends to spread 
and to shift toward larger values. Moreover, even if the TMs use many more coding triplets than strictly necessary, we see from figure
\ref{doppio_grafico} that once they approach the error threshold, the performance growth with generations slows down considerably, 
while the number of coding triplets remains practically constant. In this way, an ``historical'' factor is introduced into the evolutionary 
dynamics; namely, once the TMs have wasted their coding triplets, they  need a very large amount of time to achieve a more efficient 
usage. 

The mathematical model developed in the Methods showed that under the following hypotheses
\begin{enumerate}
\item $Q_i<1 \  \forall i$,
\item $Q_i$ is a monotonically decreasing function of $i$,
\item $g_{i+1,i} \neq 0 \ \forall i=1,\dots,M-1$,
\end{enumerate}
the mutation-selection dynamics will always decrease the fidelity rates of the evolving organisms (in our case the TMs), until they reach the error threshold $\bar{Q}$ or the highest performance class $M$. 
All of these hypotheses do hold true for our evolutionary model. Indeed the first one is implied by the definition (\ref{qualityTM}) of the fidelity rate for the TMs, that was also used to calculate the extinction times. 
From (\ref{qualityTM}) and from the fact that the performance and the number of coding triplets are positively correlated it follows that, on average, the fidelity rate decreases 
while the performance increases. The second hypothesis corresponds to the deterministic limit of this effect. 

The third hypothesis states that it is always possible to increase by one the performance  through mutations. 
From the simulations we see (figure \ref{doppio_grafico}) that the performance grows almost linearly with the generations until approaching the critical number of coding triplets, thus supporting this hypothesis. 

From a theoretical point of view, a performance increase of one can be obtained 
in our model in the following way. Let us suppose that the TM stops before reaching the $4000$ maximum time steps (entering into the 
halt state). Let $d$ be the distance on the output tape between the 
head position after the TM stopped and the nearest cell that would improve the performance score if its value would be changed.
The TM can then increase its performance by one by adding $d$ further coding triplets that move the machine head on the desired cell
(without altering the intermediate cells) and change its value. What it is important is that the probability for this process to happen is not
exceedingly small, so that performance increases can be observed within the generation range. 
The above mechanism does not work for non-halting machines. Indeed, since the maximum observed number of coding triplets is less than $600$, non-halting machines have to use several times some subset of them. If we introduce a mutation inside a 
coding triplet belonging to this subset, with the aim of modifying the dynamics at a given time step, we cannot predict how it will affect the
earlier dynamics (notice that, in the previous case, we are sure that the triplet calling the Halt state is executed only once). 
So, in the non-halting case, we cannot  see any simple recipe to increase the performance. 
In general, mutations inside the above subset of coding triplets will probably result in a big change in the output tape and will be almost always discarded by selection. This leads to the fact that non-halting machines need longer times to improve their performance.
We analyzed the $3740$ best performing machines that we registered at the end of the $50000$th generation and we found that
 $2730$ stop by calling the halt state, while the remaining $1010$ stop
by exhausting the $4000$ time steps. The former TMs have a better average performance ($19.57$) compared to the latter ones ($11.95$).  We found that, on average, the distance $d$ is $5.99$ for the halting TMs and is $2.47$ for the non-halting TMs. As a reference value, the average distance of two adjacent ones on the output tape is $2.44$.      
The fact that non-halting machines have, on average, lower performance and $d$ values is consistent with the above analysis.

Since in our simulations no TM reached the maximum performance of $125$, our
deterministic model predicts that they will accumulate coding triplets until reaching the error threshold. However, this model assumes an infinite number of generations, while our simulations 
last $50000$. We can see from figure \ref{doppio_grafico}.c and  \ref{doppio_grafico}.d that, when far from the critical number of coding triplets, the TMs do indeed accumulate 
coding triplets in a steady way, that depends on the values of $\pmm$ and $\pii$. While for high values of these probabilities, $50000$ generations are enough to reach 
the critical number of coding triplets, they are not for low ones (see figures \ref{nclimit}, \ref{doppio_grafico}.c,  \ref{doppio_grafico}.d and \ref{correlation}). 
We conclude that our evolutionary model gives a working example of the dynamical behaviour predicted by the deterministic model described in the previous section. 

%Notice that in all the figures \ref{ext1} and also in \ref{corrrelation} there are points that exceed the threshold while this doesn't happen in figure \ref{nclimit} and \ref{doppio_grafico}.d. In the former , data are collected for single TMs while in the latter, data are averaged on the seeds. 

\begin{figure}[!ht]
\ifcase\nofigure
\hspace{-5mm}\includegraphics[scale=0.85, viewport= 0 0 567 183]{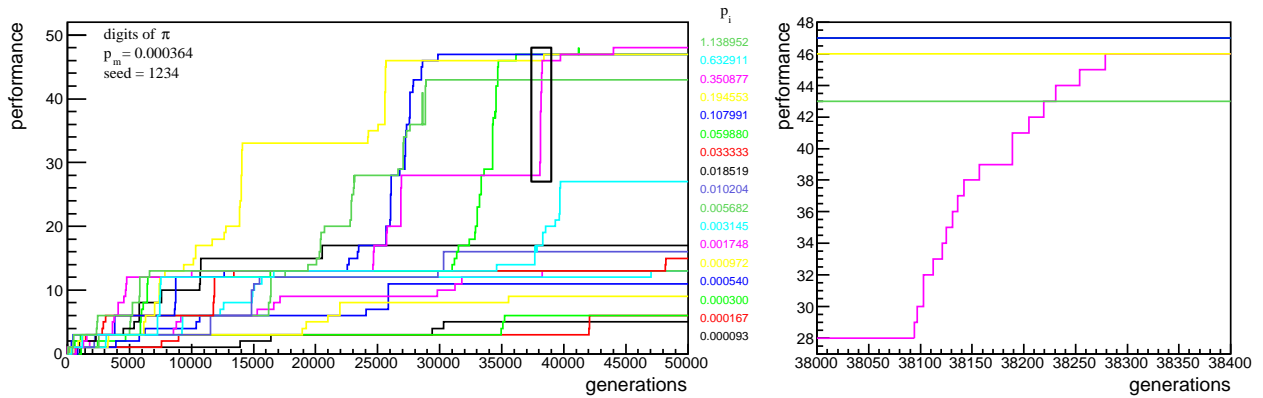}
\fi
\caption{\label{equilibria}{\bf Each coloured line shows the evolution of the performance during the generations for a single simulation.} The mutation and seed values
are shown in the upper left corner, while the state-increase rate $\pii$ is indicated on the right by the matching colour. We observe the presence of
long stasis periods alternated by short periods of fast evolution. The small black rectangle is zoomed on in the right part of the figure to show the actual jumps in the performance.}
\end{figure}

\begin{figure}[!ht]
\ifcase\nofigure
\hspace{30mm}\includegraphics[scale=0.6]{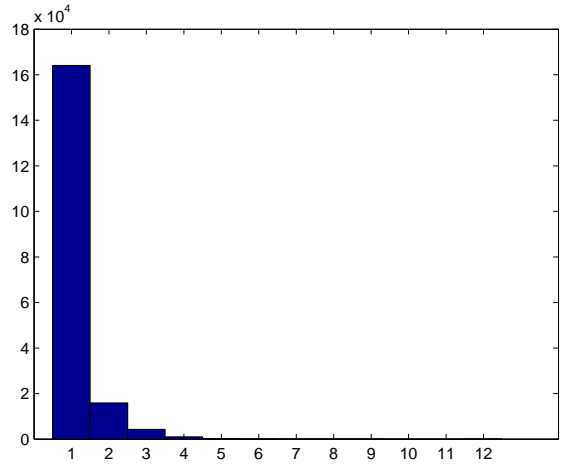}\vspace*{-5mm}
\fi
\caption{{\bf Distribution of the performance jumps versus their amplitudes.} 
This histogram shows the number of increases in the performance versus their amplitude.
\label{salti} }
\end{figure}

In figure \ref{equilibria}.a we show the time evolution of the higher performance score for all values of the state-increase probability $\pii$ corresponding to
a particular choice of the seed of the random number generator and $\pmm=3.64 \cdot 10^{-4}$. We observe the dynamical behaviour typical of punctuated equilibria \cite{Punctuated}: long periods of stasis and briefs periods of rapid evolution. The same kind of evolution 
is observed also for the other choices of the seeds, so that we present figure \ref{equilibria}.a as a representative case.   
The apparent big jumps in the performance in figure \ref{equilibria}.a, as that from $28$ to $46$ in the boxed region, are simply due to the time scale used. A zoom of the boxed region (fig. \ref{equilibria}.b) shows that this big jump is really composed by $14$ jumps of one point and $2$ jumps of two points, occurring in a relatively short number of generations. 
The same is true also for the other big performance jumps observed in figure \ref{equilibria}.a. Indeed, figure \ref{salti} shows the histogram of the number of occurrences of positive performance
jumps versus their amplitude. Performance jumps of one point (the minimum possible value) are, by far, the most common, while performance jumps larger than three are extremely infrequent. 
There is however a single, quite amazing, performance jump of $12$ that in  figure \ref{salti} is not visible due to the scale used. The mean positive performance jump in our simulations 
is $1.16$. This value seems to us sufficiently near to the minimum possible value of $1$, that we would say that the evolution of the performance in our model is essentially gradualistic. Let us stress that while the mechanism proposed in \cite{Punctuated} is based 
on a particular speciation mechanisms, in our case this dynamical behaviour is determined only by the form of the performance landscape.   

The mean performance jumps for different values of  
$\pmm$ and $\pii$ fluctuate between $1.00$ and $1.27$,
the largest value appearing for $\pmm \simeq 1.64 \cdot 10^{-3}$ and $\pii \simeq 3.51 \cdot 10^{-1}$.
Notice that this probability values do not coincide with those associated with the largest performance, 
namely $\pmm \simeq 3.64\cdot 10^{-4}$, $\pii=1.14$,
whose corresponding mean performance jump is $1.13$, that is lower than average. 

\begin{figure}[!ht]
\ifcase\nofigure
\hspace{30mm}\includegraphics[scale=0.45]{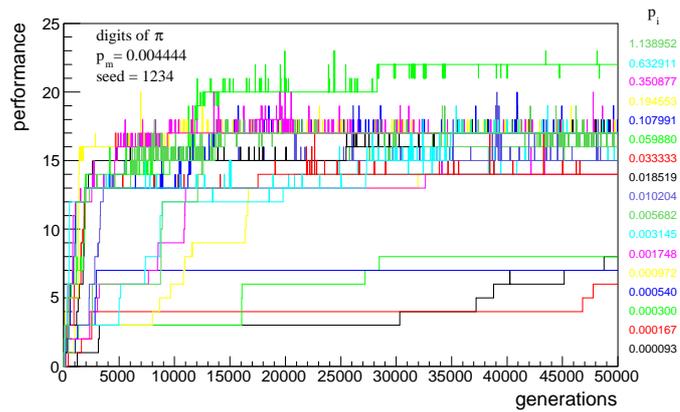} 
\fi
\caption{\label{oscillations} {\bf Performance evolution near the error threshold.} 
Here, as in figure \ref{equilibria}, we show the growth of the performance in the generations but for the much higher 
value of the mutation probability $\pmm=0.0044$.
For this value of $\pmm$ and certain values of $\pii$, TMs reach the error threshold. From there on, a typical oscillatory 
pattern emerges.}
\end{figure}

For the value of the mutation probability considered in figure \ref{equilibria}, the TMs stays far from the error threshold and we see the typical increasing trend of performance with 
generations. In figure \ref{oscillations} we present the same graph but for a much higher value of the mutation probability $\pmm$. In this case, for some values of $\pii$, the TMs do indeed 
reach the error threshold. From that moment on a typical oscillatory behaviour around a base performance value emerges. There are many performance increases followed by a rapid extinction and also performance decreases followed in a short time by back mutations.

\section{Discussion} \label{conclusions}
In this paper we studied, through computer simulations and mathematical modeling, the dynamics of an evolutionary model for Turing machines. 
In the mathematical models, by imposing suitable hypotheses on the impact of mutations on the performance landscape, we were able to compute the value of the error threshold and the expected extinction times for Turing machines versus the mutation rate. The agreement between theoretical and simulation data (see fig. \ref{ext1}) prove that the hypotheses we made are accurate
for our model. Our main finding is that evolution pushes the TMs towards the error threshold. Again, we substantiated this finding through mathematical analysis, by showing that this behaviour is
due to the mutation and selection mechanisms used and on some further hypotheses related to the structure of the performance landscape. Consequently, the question of the similarity between TMs and biological organisms is irrelevant to address the problem of the biological relevance of this finding. What is really relevant is the biological plausibility of the mutation-selection mechanisms and of the hypotheses employed. Let us stress that, despite this fact, our model still has to be considered as a toy model of evolution, so that it contains simplifying (hence necessarily unrealistic) hypotheses that makes it mathematically affordable. Nevertheless, toy models can give valuable suggestions on mechanisms working also in the full, non-simplified system of which they are approximations.
In the following we will try to discuss the hypotheses we made from a biological point of view. 

The main and most relevant approximation is to consider a unique and fixed performance landscape. In nature, the existence of different ecological niches, the changing environment and the coevolution with other species give raise to multiple and ever changing fitness landscapes. While it is difficult to estimate how this approximation influences our conclusion, it is worth noticing that
an ever changing performance landscape makes a perfectly fit organism substantially unattainable. By ruling out one of the possible end points of evolution, this fact could reinforce our results.

% The main finding is the general existence of an error threshold for ``organisms''
% subjected to mutation and selection mechanisms, and that these mechanisms push the population 
% toward the error threshold.

%This is consequence of the mutation and selection mechanisms used, and, to a minor extent, on}
%The mathematical models introduced here were used to show that one of the features that we observe in our model, namely that the mutation-selection dynamics pushes the TMs toward the error threshold, depends only on the mutation and selection mechanisms used and on 
% some further hypotheses related to the structure of the performance landscape.   

% Let us stress that the fact that we observe this behaviour in our evolutionary model is due simply to the fact that the TMs do indeed satisfy these hypotheses. Consequently, the question of the 
% similarity between TMs and biological 
% organisms is irrelevant to address the problem of the biological relevance of this finding. What is really relevant is the biological plausibility of the mutation-selection mechanisms and of the % hypotheses employed. So, in the following we will discuss this issue.   

The deterministic mutation-selection model that we used is completely specified by the selection mechanism and by the choices of the parameters $Q_j$, $j=1,\dots,M$, $g_{ij}$, $i,j=1,\dots,M$ and $f$. The values of $Q_j$ and $g_{ij}$ are related to the mutation mechanisms and to the genotype $\to$ performance mapping, while $f$ specifies, through the tournament selection, the performance $\to$ fitness mapping. Let us first discuss the selection mechanism and the choice of $f$. 
First of all, the selection mechanism keeps the total population $N$ constant (soft selection). This is a frequent assumption in population genetic models (for instance it is used in the Wright-Fisher model and in the Eigen model). From a biological point of view it translates in assuming that the population fecundity is always enough to keep it to the (constant) carrying capacity $N$ of the environment. The fact that only two individuals are compared at each generation is clearly unrealistic from a biological point of view. However, if one considers a large number of generations, virtually all individuals will have interacted through this pair interactions, so we think that a more realistic interaction mechanism would not alter the conclusions. Also the assumption that the fitness difference $f$ depends on the performance values of the two individuals only through the signum of their difference is not realistic. However, since the conclusion that the population will eventually reach the error threshold holds for any $f>0$, a more realistic choice would not change it, but only affect the population distribution in performance classes near the error threshold.

Regarding the mutation mechanism, we needed three basic assumptions:    
\begin{enumerate}
\item there is no perfect replicator,
\item the fidelity rates and the performance classes are negatively correlated,
\item the probability of improving the performance is never exceedingly small.
\end{enumerate}
The first and the third hypotheses seems to us perfectly acceptable from a biological point of view. The second assumption deserves a deeper discussion. It can be interpreted as saying that the performance of an individual is incremented mainly through the addition of coding DNA
%\footnote{
(here we use ``coding'' in the same informatic/algorithmic sense that we used for our TMs model, 
to indicate parts of the genome that influence the phenotype; the usual biological meaning would be to indicate protein coding sequences), and that this addition increases the probability of undergoing a non-neutral mutation. 

The latter statement is quite natural: if, for example, an organism increases its performance by converting a piece of junk DNA into a new gene (the metabolic cost should be kept into account to evaluate if there is a real performance increase), all the mutations that inactivate the new gene will be new non-neutral mutations. According to the first part, one has to assume 
that organism improve their adaptation more by increasing their coding DNA than by reorganizing it.

We suggested that the fact that the mutation-selection dynamics pushes the evolving organisms towards the error threshold is due to quite mild hypotheses and could be a quite robust property of evolutionary systems. Indeed, the same phenomenon appears in another artificial evolution experiment \cite{Knibbeetal2007b}, where the codification of the evolving algorithms and the mutation 
and selection mechanisms used are completely different from ours.  

At the biological level, RNA viruses have error rates (per genome per replication) near to one and have been suggested to replicate near the error threshold (see for example \cite{Eigen2000} and the references therein), in accordance with the behaviour that we observe in our model. Indeed, these organisms lack proof-reading mechanisms, so that their mutation rates per nucleotide are quite large. Moreover, the necessity to escape the immune response produces a selective pressure toward an 
high variability. Notice, however, that this latter effect is not present in our evolutionary model, since we considered a static performance landscape. 
For what concerns DNA based organisms, they also have remarkably small variations in their mutation rates\cite{Drake}.  However, 
their mutation rates are much smaller than those of RNA viruses, of the order of $1/300$ per genome per replication.
In \cite{Eigen2000} it has been suggested that these almost constant mutation rates could be due to the fact that also DNA based organisms 
do reproduce near the error threshold. Their higher fidelity rate could be explained by two factors: a larger number of neutrals in DNA sequences and the dissymmetry between the error rates of the two daughter DNA double strands (see \cite{Eigen2000}). 
While our results could encourage this explanation, the development of error correction mechanisms is not considered in our model.
Approaching the error threshold surely induces an high selective pressure on the development of error correction mechanisms. 
The short term advantage derived by an increase in the reproductive fidelity could be reinforced by the long term advantage of an 
higher evolvability (if the organisms evolvabilities are much lower near the error threshold as it happens in our model, see figure 
\ref{doppio_grafico}). Maybe, the mutation rates observed for the DNA could be due to a balance between the natural trend toward 
the error threshold, due to the mutation-selection dynamics, the need of proof-reading mechanisms and their metabolic costs. 

In this paper, we showed that some of the features observed in an artificial evolutionary model can have a much more general validity than the specific model itself.
This happens when the phenomenon under consideration is mainly due to the mutation-selection dynamics, so that it can be described through a population genetic model. 
It seems to us that this synergistic integration between artificial evolution and population genetic model should be pursued, when possible. Since the phenomena observed in an 
artificial evolutionary model can have a quite wide degree of generality, we think that they can give interesting suggestions on possible evolutionary mechanisms working also at the
biological level.

%\section*{Acknowledgments}
\section*{Funding}
This work was partially supported by the Spanish Ministerio de Ciencia e
Innovaci\'on  under grant MTM2007-67389 (with EU-FEDER support), by  
Junta de Castilla y Le\'on  (Project GR224) and by UBU-Caja de Burgos (Project K07J0I).

%%%%%%%%%%%%%%%%%%%%%%%%%%%%%%%%%%%%%%%%%%%%%%%%%%%%%%%%%%%%%%%%%%%%%%%%%%%%%%%%%%%%%%%%%%%%%%%%%%%%%

%%%%%%%%%%%%%%%%%%%%%%%%%%%%%%%%%%%%%%%%%%%%%%%%%%%%%%%%%%%%%%%%%%%%%%%%%%%%%%%%%%%%%%%%%%%%%%%%%%%%%%%%%%%
\newpage
\ifcase\nofigure
\section*{Figures}
\or
\section*{Figure Legends}
\fi

%%%%%%%%%%%%%%%%%%%%%%%%%%%%%%%%%%%%%%%%%%%

%%%%%%%%%%%%%%%%%%%%%%%%%%%%
%%%%%%%%%%%%%%%%%%%%%%%%%%%%
%%%%%%%%%%%%%%%%%%%%%%%%%%%%%
%%%%%%%%%%%%%%%%%%%%%%%%%%%%%%%%%

%%%%%%%%%%%%%%%%%%%%%%%%%%%%%%%%%%%%%%%%%%%%%%%%%%%%
%%%%%%%%%%%%%%%%%%%%%%%%%%%%%%%%%%%%%%%%%%%%%%%%
%%%%%%%%%%%%%%%%%%%%%%%%%%%%%%%%%%%%%%%%%%%%%%%%%%%%%%
%%%%%%%%%%%%%%%%%%%%%%%%%%%%%%%%%%%%%%%%%%%%%%%%%%%%%%%%%%%%%%

%%%%%%%%%%%%%%%%%%%%%%%%%%%%%%%%%
%%%%%%%%%%%%%%%%%%%%%%%%%%%%
%%%%%%%%%%%%%%%%%%%%%%%%%%%%

\end{document}